\documentclass[10pt,a4paper]{article}
\usepackage[a4paper,margin=1in]{geometry}
\usepackage{graphicx, multirow, outlines, ulem}
\usepackage[latin1]{inputenc}
\usepackage{amsmath} 
\usepackage{amsfonts}
\usepackage{amssymb}
\usepackage{caption}
\usepackage{subcaption}
\usepackage{float}
\usepackage{cite}
\usepackage{epstopdf}
\DeclareGraphicsExtensions{.pdf,.png,.jpg}
\usepackage{breqn}
\usepackage{upgreek}
\makeatother
\makeatletter
\newcommand{\mathleft}{\@fleqntrue\@mathmargin0pt}
\newcommand{\mathcenter}{\@fleqnfalse}
\usepackage{xcolor}
\usepackage[colorlinks,citecolor=red,linkcolor=blue]{hyperref}
\def\beq{\begin{equation}}
\def\eeq{\end{equation}}
\def\bea{\begin{eqnarray}}
\def\eea{\end{eqnarray}}

\begin{document}

\begin{center}
  {\Large \bf  New solutions of Isochronous potentials in terms of exceptional orthogonal polynomials in heterostructures}
\vspace{1.3cm}

{\sf Satish Yadav\footnote[1]{e-mail address:\ \ s30mux@gmail.com,\hspace{0.05cm} Satish.yadav17@bhu.ac.in},
Rahul Ghosh \footnote[2]{e-mail address:\ \ rg928@snu.edu.in,\hspace{0.05cm} },
Bhabani Prasad Mandal\footnote[3]{e-mail address:\ \ bhabani.mandal@gmail.com,\hspace{0.05cm} bhabani@bhu.ac.in}}

\bigskip 

{\it $^{1,2,3}$ Department of Physics,
Banaras Hindu University, \\ Varanasi-221005, INDIA. \\ \it $^{2}$ Dynamics Lab, Indian Institute of Technology Delhi, New Delhi 110016, India.
}

\bigskip
	\noindent {\bf Abstract}
\end{center}

Point canonical transformation (PCT) has been used to find out new exactly solvable potentials in the position-dependent mass (PDM) framework. We solve $1$-D Schr\"{o}dinger equation in the PDM framework by considering two different fairly generic position-dependent masses $(i) M(x)=\lambda g'(x)$ and $(ii)  M(x) = c \left( {g'(x)} \right)^\nu $, \ $\nu =\frac{2\eta}{2\eta+1},$ with $\eta= 0,1,2\cdots $. In the first case, we find new exactly solvable potentials that depend on an integer parameter $m$, and the corresponding solutions are written in terms of $X_m$-Laguerre polynomials. In the latter case, we obtain a new one parameter $(\nu)$ family of isochronous solvable potentials whose bound states are written in terms of $X_m$-Laguerre polynomials. Further, we show that the new potentials are shape invariant by using the supersymmetric approach in the framework of PDM. 

\medskip
\vspace{1in}

\section{Introduction}    

In quantum mechanics, the Schr\"{o}dinger equation for a few systems is solved exactly though the traditional means \cite{parfitt02_two}, the factorization methods of supersymmetric quantum mechanics (SUSY) \cite{cooper95,bag00,fern10,junker12_supersymmetric,gango17,yadav22}, Lie algebraic techniques \cite{wu90,engle91,bag22},  reduction approach (to a hypergeometric equation \cite{manning35_exact} or to a confluent Heun equation \cite{downing13_solution}), Nikiforov-Uvarov (NU) method \cite{hass13,ghosh22_solving} etc. Searching for analytically solvable potentials, whose energy spectrum are completely known, constitute an important field to study the low dimensional structures e.g. quantum well, quantum dot \cite{harrison16_quantum}, specially the linear and nonlinear optical properties of the same \cite{guo93_nonlinear,el14_linear,vahdani14_effect}. On the other hand, the scenario of non-constant mass particularly position-dependent mass (PDM), although initially pursued in condensed matter physics problems, has become an active area of research in many disciplines of physics. The significant advancement in crystallographic growth techniques that enable the manufacture of non-uniform semiconductor specimens with abrupt heterojunctions is what has drawn the attention to the PDM approach. This kind of  spatial dependencies have provided useful insights into new classes of phenomena such as  the transport proprieties in semiconductors \cite{harrison16_quantum,bastard90_wave}, compositionally graded crystals \cite{geller93}, quantum dots \cite{serr97}, liquid crystals \cite{barr97}, and extended systems governed by superintegrability \cite{miller13a,marquette08}. The concept of PDM is also used for generating the pseudo-potentials which have an essential  computational advantage in quantum Monte Carlo method \cite{foulkes01_quantum,martin20_electronic}. On aside earlier studies on fundamental length scale (to remove divergences in field theories) show that any modification to the underlying space or to the canonical commutation relations (i.e Heisenberg uncertainty relation) \cite{kempf95,kempf09} can result to a Schr\"{o}dinger equation with a position-dependent mass \cite{quesne04deformed,filho11_displacemnt}. To study these inhomogeneous materials we need to have an effective way to solve the PDM Schr\"{o}dinger equation (PDMSE) which we will address in this article. Before going further it needs to be emphasized that in any scheme for the same leads to non-commutativity of the momentum and mass-operators and it results in an ordering ambiguity in the kinetic energy (KE) representation (see, e.g., \cite{must06,cavalcante7_form}). To avoid this we will use the Ben-Daniel and Duke model of PDMSE \cite{bend66} for our study.

The bound state solutions of the Schr\"{o}dinger equation are generally associated with some classical orthogonal polynomials such as Hermite polynomials, Laguerre polynomials, Legendre polynomials, etc. In 2009, new families of orthogonal polynomials connected to their classical counterparts were introduced and termed exceptional orthogonal polynomials (EOPs)\cite{gomez09,gomez10,gomez12}. These polynomials form an orthogonal and complete set of polynomials with a positive weight function even if their sequence starts with a degree $n\ge 1$. After this remarkable discovery, the quantum mechanical exploration of exactly solvable systems has been boosted, particularly those whose solutions are written in terms of EOPs. These are known as rational extensions of usual quantum systems. Over the past decade, these advancements have significantly expanded the list of exactly solvable potentials.\cite{yadav2019rationally,yadav2022one,kumari2017class,kumari2018class,kumari2016scattering,yadav2013scattering,yadav2013scattering1,dutta2010conditionally}.


In this article, we consider the PCT approach to construct the new exactly solvable potentials in a PDM background. We study the two cases, in the first case, we consider $M(x)= \lambda g'(x)$ which was also considered \cite{midya09} to obtain the solution of PDMSE in terms of $X_1$-Laguerre polynomials. We find new exactly solvable potentials and the eigenstates associated with these potentials are written in terms of $Xm$-Laguerre EOPs. It is important to note that numerous research have already been done on the intriguing problem of how classical and quantum "generalised harmonicity" relate to one another to which end we encounter with isochronous potentials \cite{stillinger89_pseudoharmonic,dorignac05_quantum}. Stillinger and Stillinger \cite{stillinger89_pseudoharmonic} have suggested a two-parameter family of isochronous potentials, interpolating between the harmonic oscillator and the asymmetric parabolic well, while Dorignac has explored the quantum spectrum of the same which comes as equispaced \cite{dorignac05_quantum}. In the second case, we propose a general form of $M(x) = c \left( {g'(x)} \right)^\nu, \ \nu =\frac{2\eta}{2\eta+1},$ with $\eta= 0,1,2\cdots  $ and obtain a one-parameter family of isochronous potentials that are exactly solvable and asymmetric in nature. Solutions are constructed in terms of $X_m$-Laguerre EOPs. It is worth adding that a similar study has already been undertaken for different choices of $M(x)$\cite{bagchi2005_new,midya09,quesne08a}.
Furthermore, by using the supersymmetric approach in quantum mechanics in the context of a PDEM framework, we show that these exactly solvable potentials are shape-invariant.

The organization of this paper is as follows: in section $2$ we briefly introduce the position-dependent mass  Schr\"{o}dinger equation and deformed algebra. In section $3$ we have generated two new exactly solvable potentials by using the point canonical transformation approach in the position-dependent effective mass background whose wavefunctions are written in terms of $X_m$  Laguerre orthogonal polynomials. In section $4$ we have shown that these new exactly solvable potentials are shape invariant via using SUSY in quantum mechanics in PDM background. Finally, in section $5$ we summarize the result.

\section{PDM Schr\"{o}dinger Equation and Deformed Algebra} \label{sec: PDMSE and deformded algebra}

To avoid the noncommutativity, we start with the von Roos prescription of the two-parameter formulation of an effective-mass kinetic energy (KE) operator \cite{vonr83}, which has an inbuilt Hermiticity (which also yields different plausible special cases), reads as 
\begin{equation} 
	\hat{T} = \frac{1}{4}[\mathcal{M}^r (x) p \mathcal{M}^s (x) p \mathcal{M}^t (x) +\mathcal{M}^t (x) p {\mathcal{M}}^s (x) p\mathcal{M}^r (x)]
\end{equation}
where $r$, $s$ and $t$ constitute a set of ambiguity parameters which follow $ r +s +t = -1 $. Using the above $\hat{T}$, we can write the Hamiltonian for PDMSE as follows
\begin{equation} \label{vonroos}
	H = \Big[ \frac{1}{4}[\mathcal{M}^r (x) p \mathcal{M}^s (x) p \mathcal{M}^t (x) + \mathcal{M}^t (x) p \mathcal{M}^s (x) p \mathcal{M}^r (x)]  +V(x) \Big] ,
\end{equation}
where $V(x)$ is the external potential. Of the frequently used Hermitian Hamiltonians, a few possible choices of the ambiguity parameters, have been explored in  the literature for that see \cite{bag20b} and references therein.


Adopting units such that $\hbar = 2\mathcal{M}_0 = 1$ the PDMSE corresponding to the Hamiltonian in (\ref{vonroos}) can be projected as 
\begin{flalign} \label{pdminto_VonR H2}
	\Big[-\frac{1}{2} \Big( M^r (x) \frac{d}{dx} M^s (x) \frac{d}{dx} M^t (x) + M^t (x) \frac{d}{dx} M^s (x) \frac{d}{dx} M^r (x) \Big) +V(x) \Big] \psi(x)= E \psi(x) 
\end{flalign}
where $M(x)$ is defined by $\mathcal{M}(x)=\mathcal{M}_0 M(x)$ and is dimensionless. On setting $ M(x)= f^{-2}(x) $ and $ \quad f(x)=1+\mathfrak{g}(x) $ \cite{bag05}, where $f(x)$ is some positive-definite function and $\mathfrak{g}(x) = 0$ corresponds to the constant-mass case, equation (\ref{pdminto_VonR H2}) becomes
\begin{flalign} \label{deformedSE1}
	\Big[ -\frac{1}{2} \Big( f^r (x) \frac{d}{dx} f^s (x) \frac{d}{dx} f^t (x) + f^t (x) \frac{d}{dx} f^s (x) \frac{d}{dx} f^r (x) \Big) +V(x) \Big] \psi(x)= E \psi(x) 
\end{flalign}
with $r+s+t=2$. We can get rid of the ambiguity parameters $r$, $s$, $t$ (denoted collectively by {\it a}) in the kinetic energy term by transferring them to the effective potential energy of the variable-mass system. Then using the result
\begin{gather} \label{f tranfromation}
	f^r \frac{d}{dx} f^s  \frac{d}{dx} f^t + f^t \frac{d}{dx} f^s  \frac{d}{dx} f^r 
	=2 \sqrt{f}\frac{d}{dx}f\frac{d}{dx}\sqrt{f}-(1-r-t)f f^{''}-2(\frac{1}{2}-r)(\frac{1}{2}-t)f^{'2}
\end{gather}
where the prime denotes derivative with respect to $x$ and the positive definiteness of $f$ is explicitly assumed, equation (\ref{deformedSE1}) acquires the form
\begin{align}  \label{deformed H1}
	H\psi(x)= \big[ -\Big( \sqrt{f(x)}\frac{d}{dx}\sqrt{f(x)} \Big)^2+ V_{eff}(x) \Big]\psi(x)= E \psi(x)
\end{align}
in which the effective potential
\begin{gather}  \label{deformed Veff}
	V_{eff}(x)=V(x) +  \frac{1}{2}(1-r-t) f(x) f''(x)+ (\frac{1}{2}-r)(\frac{1}{2}-t) f^{'2}(x)
\end{gather}

The PDMSE (\ref{deformed H1}) may now be reinterpreted as the deformed Schr\"{o}dinger equation as given by
\begin{align}  \label{deformed H2}
	H\psi(x)= \big[ \pi^2 + V_{eff}(x) \Big]\psi(x)= E \psi(x)
\end{align}
where the standard momentum operator $p=-i\frac{d}{dx}$ is now modified by the following deformed quantity defined by
$ \pi = - i\sqrt{f(x)}\frac{d}{dx}\sqrt{f(x)} $ and 
we thus observe that with respect to $\pi$ the conventional commutation relation $[x, p]= i~\hbar$ is modified to $ [x,\pi]= i~\hbar f(x) $ where $f(x)$ acts as a deforming function\cite{kempf95,bag05,izadparast20}. This type of deformation also has been extensively studied in different areas of physics such as in statistical physics (knowing as $q-$deformed algebra)\cite{tuszynski93_statistical,lavagno00_themostat}, in optics \cite{braidotti17_generalized,conti14_quantum} as well as black hole physics \cite{bosso21_deformed,bera22_quantum}. In the next section, we will use the PCT and EOPs to obtain a class of exactly solvable potentials and their eigen spectrum. 

\section{Generation of new potential by PCT approach}
Following the procedure outlined in the previous section the PDMSE in Eq. (\ref{pdminto_VonR H2}) can be written as 
\begin{equation}
	H \psi(x)=\Big[ - \frac{d}{dx}\frac{1}{M(x)} \frac{d}{dx}+V_{eff} \Big] \psi(x) = E \psi(x)
	\label{1}
\end{equation}
where  the effective potential is
\begin{equation}
	V_{eff}(x)=V(x)+\frac{s+1}{2} \frac{M''}{M^2} -[r(r+s+1)+s+1]\frac{{M'}^2}{M^3}
	\label{2}
\end{equation}
$V(x)$ is the applied potential in the system.  We will follow the choice of BenDaniel and Duke \cite{bend66 }, i.e. $r=t=0 $ and $s=-1$ for which the effective and applied potentials become identical.
We use the solution $\psi(x)$ in the form given by \cite{bag05,quesne08a,midya09}
\begin{equation} \label{psi f(x)F(g(x))}
	\psi(x)=f(x) F(g(x))
\end{equation}
where $F(g(x))$ satisfies second order differential equation
\begin{equation}\label{3}
	\frac{d^2 F}{dg^2}+Q(g)\frac{dF}{dg}+R(g)F=0
\end{equation}

Now substituting this $\psi(x)$ in  Eq. (\ref{1}) we get
\begin{equation}
	\frac{d^2 F}{{dg}^2}+\left(\frac{g''}{{g'}^2}+\frac{2f'}{fg'}-\frac{M'}{Mg'}\right) \frac{dF}{dg} + \Big(\frac{f''}{f {g'}^2} - \frac{M' f'}{Mf{g'}^2} +(E-V_{eff})\frac{M}{{g'}^2} \Big)F=0
\end{equation}

On comparison we get
\begin{equation}
	Q(g(x))=\Big(\frac{g''}{{g'}^2}+\frac{2f'}{fg'}-\frac{M'}{Mg'} \Big) \label{Qintermsoffg}
\end{equation}
\begin{equation}
	R(g(x))= \left(\frac{f''}{f {g'}^2} - \frac{M' f'}{Mf{g'}^2} +(E-V_{eff})\frac{M}{{g'}^2} \right)    \label{Rintermsoffg}
\end{equation}

Eq. (\ref{Qintermsoffg}) admits the solution \cite{quesne08a}
\begin{equation}\label{fx}
	f(x) \propto \sqrt{\frac{M}{g'}} e^{\frac{1}{2}\int^{g(x)} Q(t) dt}
\end{equation}

Putting it in (\ref{Rintermsoffg})
we get 
\begin{equation} \label{EVeff1}
	E-V_{eff}=\frac{g'''}{2Mg'}-\frac{3}{4M} \left(\frac{g''}{g'}\right)^2 +\frac{{g'}^2}{M} \left( R-\frac{1}{2} \frac{dQ}{dg}-\frac{Q^2}{4}\right)- \frac{M''}{2M^2} +\frac{3{M'}^2}{4M^3} 
\end{equation}

To ensure the presence of an energy-like term on the right-hand side of the above equation, which will compensate for the term $E$ on its left-hand side, we can introduce a suitable relation between $M(x)$ and $g(x)$. This will give rise to an effective potential $V_{eff}(x)$ and the solution of the corresponding Schr\"{o}dinger equation will be well-behaved. Three different choices of $M(x) \propto \left( {g'(x)} \right)^\mu$ with $\mu=-1,1,2$ have been used extensively in various applications of the PCT approach in the PDM context\cite{alhaidari2002_solutions,bagchi2005_new}. In this article, we consider a new type of position-dependent mass 
\begin{align}
	M(x) = c \left( {g'(x)} \right)^\nu  
\end{align}
where $c$ is a proportionality constant, $\nu = {\frac{2 \, \eta}{2 \, \eta + 1 }} $ and $\eta = 0, 1, 2\cdots $ along with $M(x)=\lambda g'(x)$  
\subsection{Case 1}
For choice of $M(x)=\lambda g'(x)$, Eq.(\ref{EVeff1}) becomes 
\begin{equation}\label{ModEVeffM}
	E-V_{eff}=\frac{g'}{\lambda}\Big( R-\frac{1}{2} \frac{dQ}{dg}-\frac{Q^2}{4}\Big).
\end{equation}
We further consider the differential equation for $X_m$-Laguerre polynomials and compare it with equation (\ref{3}) to get 
\begin{equation}
	\begin{gathered}
		Q=\frac{1}{g}\left((\alpha+1-g)-2g\frac{L_{m-1}^\alpha(-g)}{L_{m}^{\alpha-1}(-g)}\right) , \quad \text{and }  R=\frac{1}{g}\left(n-2\alpha\frac{L_{m-1}^\alpha(-g)}{L_{m}^{\alpha-1}(-g)}\right) .
	\end{gathered}
\end{equation}
where $F(g)\propto \hat{L}_{n,m}^{(\alpha)}(g)$.\\
For these values of $Q(g(x))$ and $R(g(x))$ equation (\ref{ModEVeffM}) reads
\begin{align} \label{XmEVeffM}
 E-V_{eff}&=\frac{g'}{\lambda g}\left(n-\left(m-\frac{1}{2}\right)+\frac{\alpha}{2}+\frac{m}{\alpha}\right)-\frac{g'}{4 \lambda}-\frac{g'}{4g^2\lambda}\left(\alpha^2-1\right)-\frac{2g'{L_{m}^{\alpha}(-g)}^2}{\lambda\left({L_{m}^{\alpha-1}(-g)}\right)^2}\nonumber\\&\quad+\frac{g'{L_{m-2}^{\alpha+1}(-g)}}{\lambda\left({L_{m}^{\alpha-1}(-g)}\right)}-\frac{g'{L_{m-1}^{\alpha}(-g)}}{\lambda\left({L_{m}^{\alpha-1}(-g)}\right)}-\frac{g'{L_{m-1}^{\alpha+1}(-g)}}{\alpha \lambda\left({L_{m}^{\alpha-1}(-g)}\right)}+\frac{g'{L_{m-1}^{\alpha+1}(-g)}}{\lambda\left({L_{m}^{\alpha-1}(-g)}\right)}
\end{align}
We observe that choice  $\frac{g'}{\lambda g}=C$ will provide a constant term on the right-hand side to compensate for the energy term on the left-hand side. To get increasing eigenvalues for successive n-values, $C$ must be restricted to positive. The solution of the choice $\frac{g'}{\lambda g}=C$ leads
\begin{equation}\label{mxgxM}
	g(x)=e^{-bx} ~~\text{and}~~ M(x)=e^{-bx},~~~~~  -\infty < x <\infty
\end{equation}
where we have considered $C=b^2$ and $C\lambda=-b,~b > 0$.\\
Substituting Eq.(\ref{mxgxM}) in (\ref{XmEVeffM}) we obtain the effective potential and energy eigenvalues 
\begin{align}\label{Vs}
 V_{eff}&=\frac{b^2}{4}\left( e^{bx}\left(\alpha^2-1\right)+ e^{-bx}\right) + b^2\Bigg(\frac{2(L_{m-1}^{\alpha}(-e^{-bx}))^2}{\left({L_{m}^{\alpha-1}(-e^{-bx})}\right)^2}-\frac{{L_{m-2}^{\alpha+1}(-e^{-bx})}}{{L_{m}^{\alpha-1}(-e^{-bx})}}\nonumber\\&\quad + \frac{{L_{m-1}^{\alpha+1}(-e^{-bx})}}{\alpha {L_{m}^{\alpha-1}(-e^{-bx})}}+\frac{{L_{m-1}^{\alpha}(-e^{-bx})}}{{L_{m}^{\alpha-1}(-e^{-bx})}}-\frac{{L_{m-1}^{\alpha+1}(-e^{-bx})}}{{L_{m}^{\alpha-1}(-e^{-bx})}}\Bigg) e^{-bx}+V^I_c
\end{align} 

   \begin{equation}
	E_n^m=b^2\left(n+\frac{\alpha+1}{2}+\frac{m}{\alpha}\right)+V^I_c,  ~~~n=0,1,2,...
\end{equation}
where $V_c^I$ is  an arbitrary constant.
The eigenfunctions are obtained using Eqs.(\ref{psi f(x)F(g(x))}) and (\ref{fx})
\begin{equation}
	\psi_n^m(x) = \mathcal{N}\frac{\exp{[-\frac{1}{2}((\alpha+1)bx+e^{-bx})]}}{L_{m}^{\alpha-1}(-e^{-bx})}\hat{L}_{n+m,m}^\alpha(e^{-bx})  ~~~n=0,1,2,...
\end{equation}
where $\mathcal{N}$ is the normalization constant given by
\begin{equation*}
    \mathcal{N}=\left(\frac{b~n!}{(n+m+\alpha)\Gamma (n+\alpha)}\right)^{\frac{1}{2}}
\end{equation*}
We would like to point out that $m=1$ case was done in \cite{midya09} our general result for $X_m$ case reproduces the result of \cite{midya09} as 

\begin{equation}
V_{eff}=\frac{b^2}{4}\left( e^{bx}(\alpha^2-1)+ e^{-bx}+\frac{4}{\alpha(1+\alpha e^{bx})}+\frac{8}{(1+\alpha e^{bx})^2}\right)+V^I_c
\end{equation} 

    \begin{equation}
	E_n^1=b^2\left(n+\frac{\alpha+1}{2}+\frac{1}{\alpha}\right)+V^I_c,  ~~~n=0,1,2,...
\end{equation}

    \begin{equation}
	\psi_n^1(x) = \mathcal{N}\frac{\exp{[-\frac{1}{2}((\alpha+1)bx+e^{-bx})]}}{\alpha+e^{-bx}}\hat{L}_{n+1}^\alpha(e^{-bx})  ~~~n=0,1,2,...
\end{equation}

\begin{figure}[H]
	\centering
	\begin{subfigure}[b]{0.4\textwidth}
		\centering
		\includegraphics[width=6.1cm]{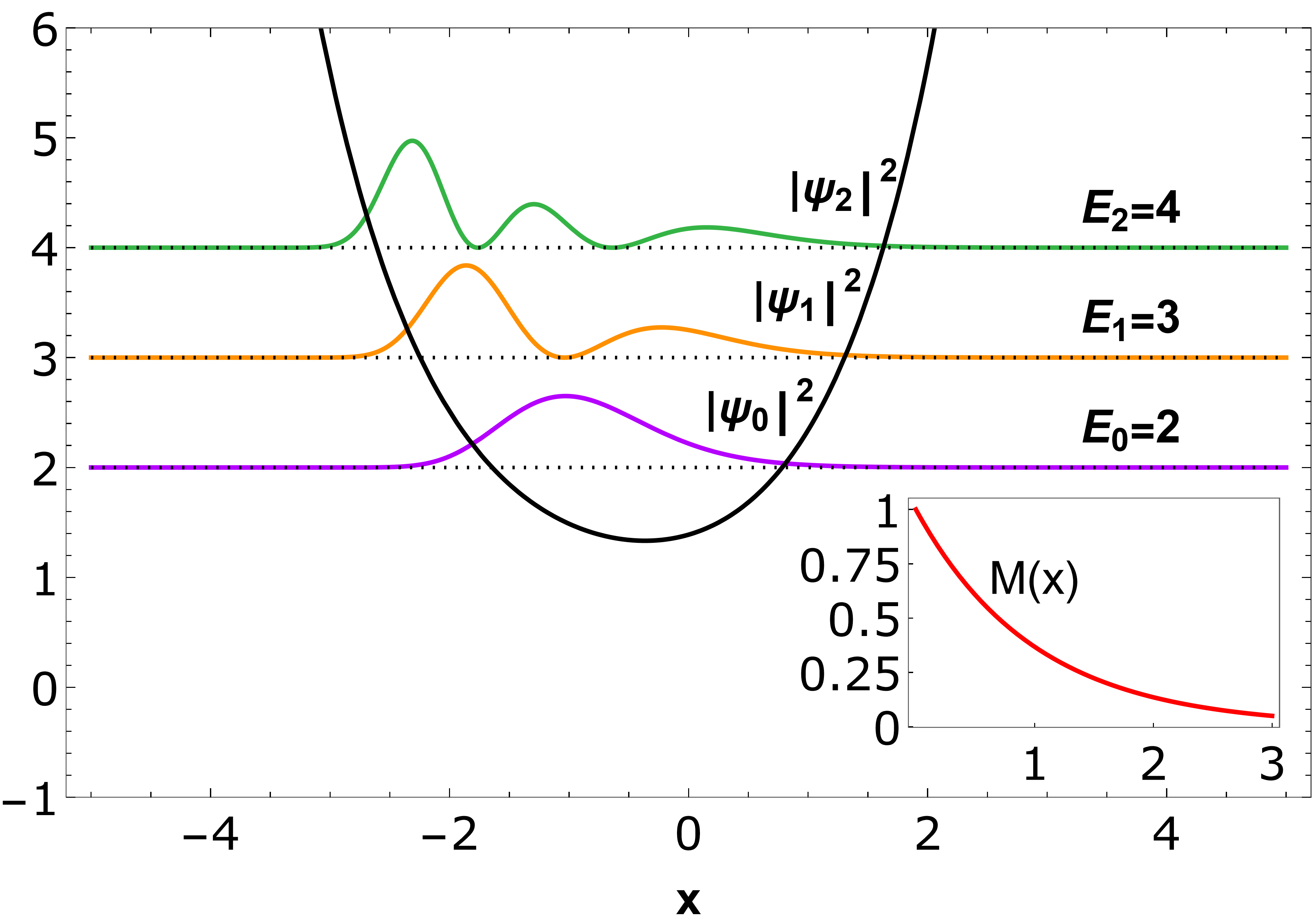}
		\caption{$m=1$}
	\end{subfigure}
	\begin{subfigure}[b]{0.4\textwidth}
		\centering
		\includegraphics[width=6.1cm]{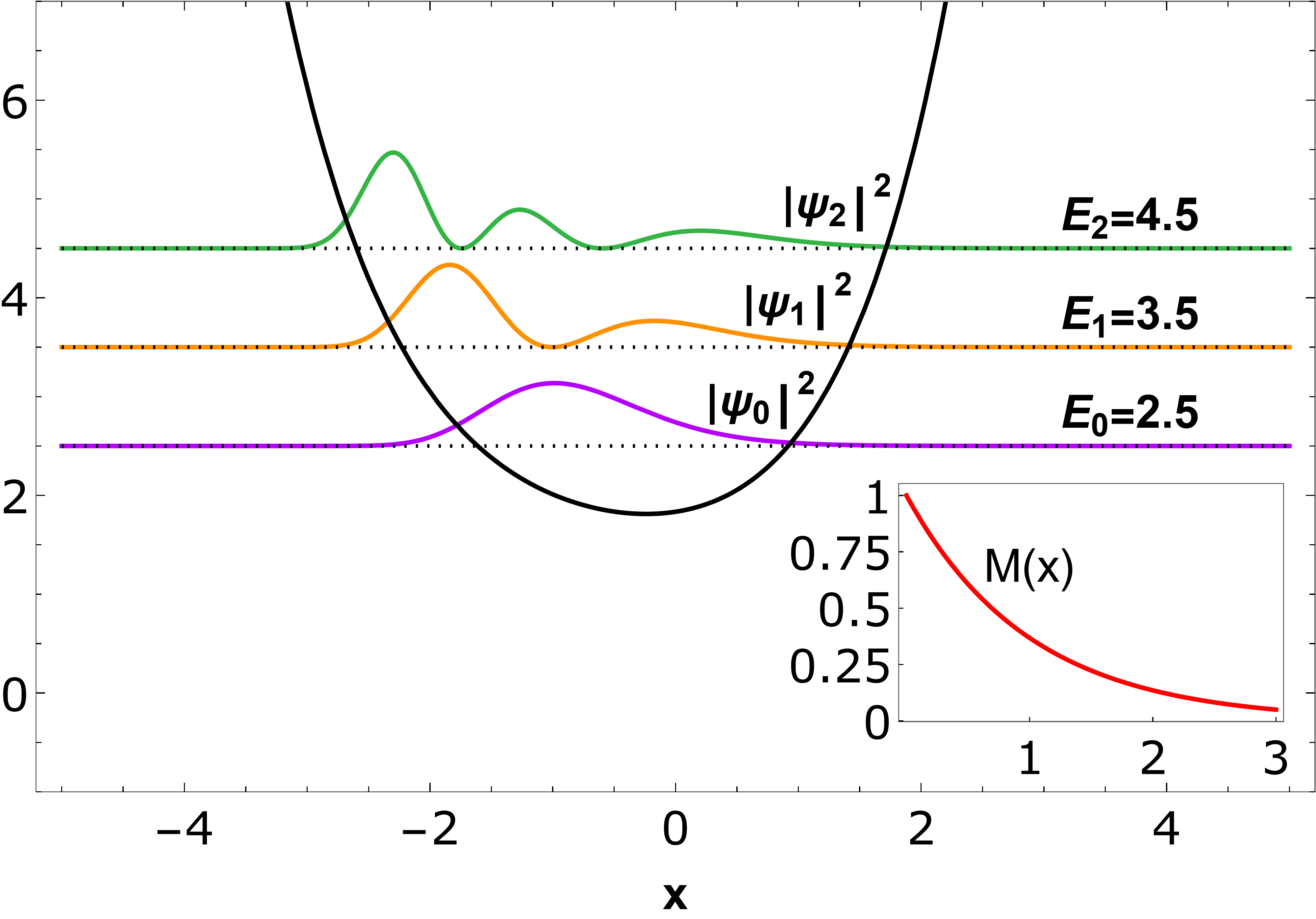}
		\caption{$m=2$}
	\end{subfigure}
	\begin{subfigure}[b]{0.4\textwidth}
		\centering
		\includegraphics[width=6.1cm]{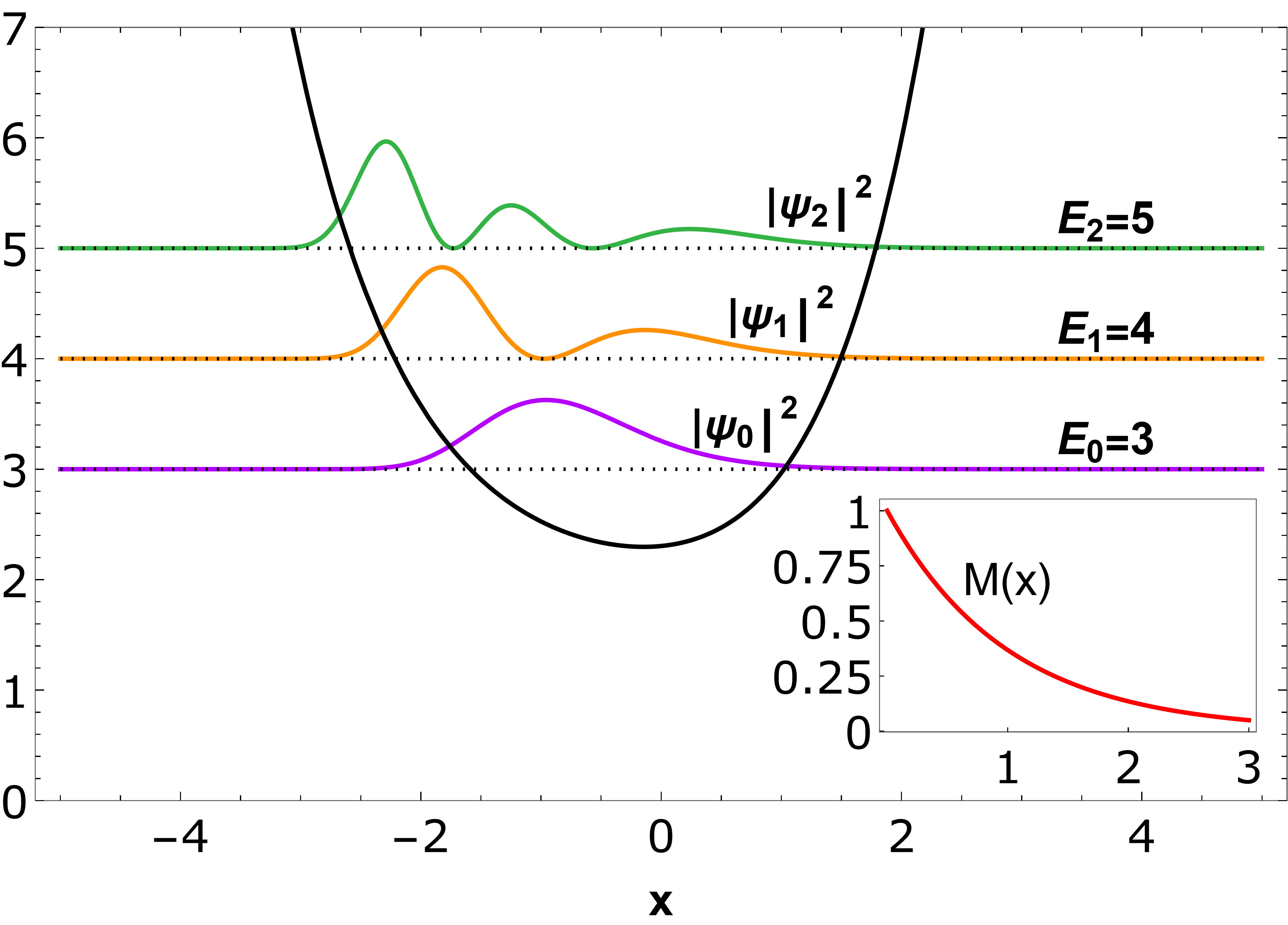}
		\caption{$m=3$}
	\end{subfigure}
	\begin{subfigure}[b]{0.4\textwidth}
		\centering
		\includegraphics[width=6.1cm]{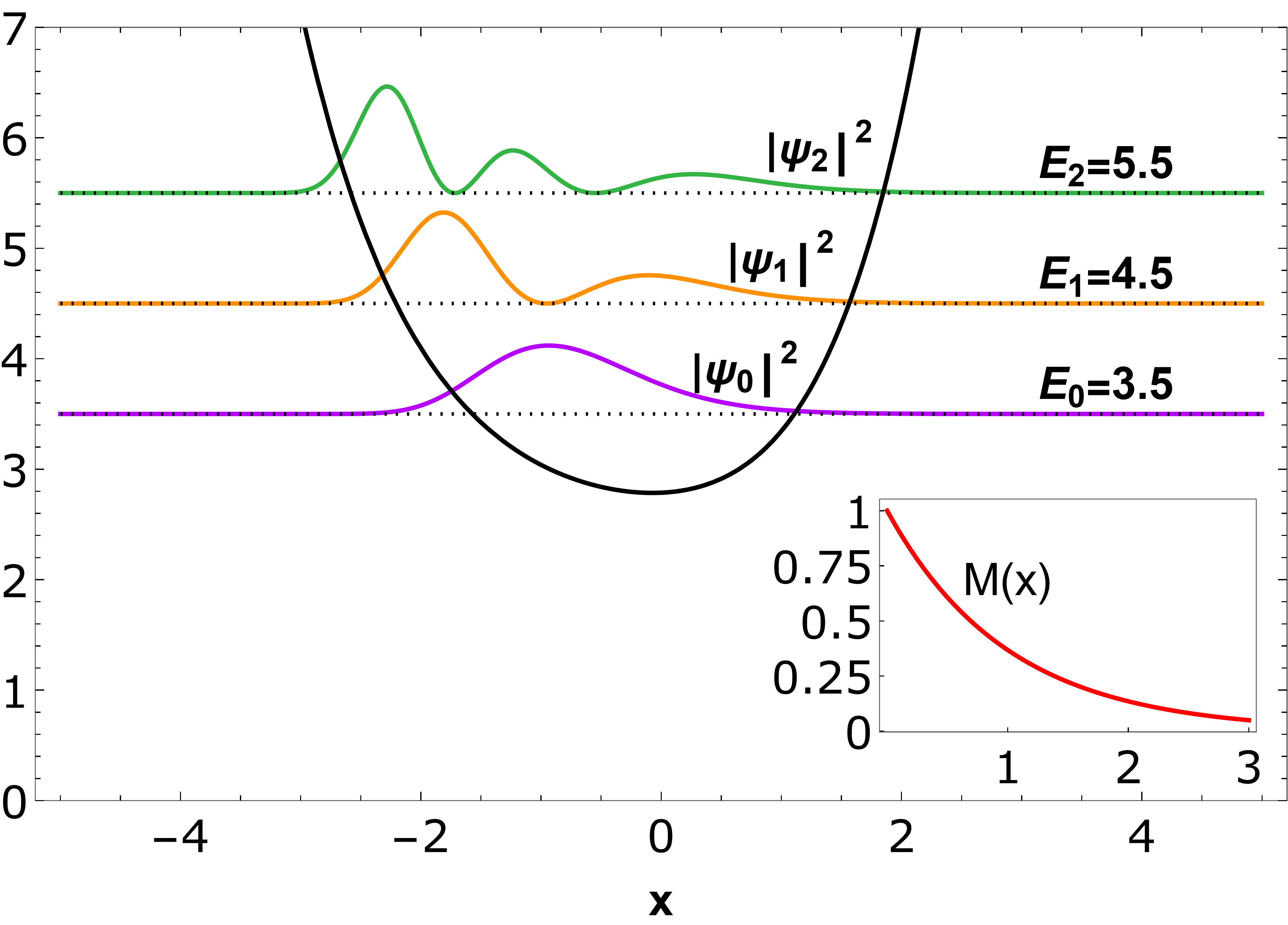}
		\caption{$m=4$}
	\end{subfigure}
	\caption{Plot of the potential $V_{eff}$ given in Eq. (\ref{Vs}) for different $m$ values, square of first three bound state wavefunctions ${|\psi_0|}^2$ (purple line), ${|\psi_1|}^2$ (orange line),  ${|\psi_2|}^2$ (green line) corresponding to different $m$ values, for mass function $M(x)$ (red line) given in Eq. (\ref{mxgxM}). We have consider here $b=1$, $\alpha=2$ }
\end{figure}

\subsection{Case 2}
In this case, we consider a new type of position-dependent mass $M(x) = c \left( {g'(x)} \right)^\nu,\nu = {\frac{2 \, \eta}{2 \, \eta + 1 }} $ and $\eta = 0, 1, 2\cdots$. For this choice Eq. (\ref{EVeff1}) becomes
\begin{align}  \label{ModEVeff}
	E-V_{eff}& = \frac{g'''}{2c{g'}^{(\nu+1)}}-\frac{3{g''}^2}{4c{g'}^{(\nu+2)}}+\frac{1}{c{g'}^{(\nu-2)}}\Big( R-\frac{1}{2} \frac{dQ}{dg}-\frac{Q^2}{4}\Big) \nonumber \\&\quad - \frac{(\nu^2-\nu){g''}^2}{2c{g'}^{(\nu+2)}} - \frac{\nu g'''}{2c{g'}^{(\nu+1)}} + \frac{3\nu^2 {g''}^2}{4c{g'}^{(\nu+2)}}
\end{align}
We further consider the differential equation for $X_m$-Laguerre polynomials with equation (\ref{3}) to get
\begin{equation}
	\begin{gathered}
		Q=\frac{1}{g}\left((\alpha+1-g)-2g\frac{L_{m-1}^\alpha(-g)}{L_{m}^{\alpha-1}(-g)}\right) , \quad \text{and }  R=\frac{1}{g}\left(n-2\alpha\frac{L_{m-1}^\alpha(-g)}{L_{m}^{\alpha-1}(-g)}\right) .
	\end{gathered}
\end{equation}
where $F(g)\propto \hat{L}_{n,m}^{(\alpha)}(g)$.\\
For these values of $Q(g(x))$ and $R(g(x))$ equation (\ref{ModEVeff}) becomes
\begin{align} \label{XmEVeff}
 E-V_{eff}&=\frac{{g'}^{(2-\nu)}}{gc}\left(n-\left(m-\frac{1}{2}\right)+\frac{\alpha}{2}+\frac{m}{\alpha}\right)-\frac{{g'}^{(2-\nu)}}{4c}-\frac{{g'}^{(2-\nu)}}{4g^2c}\left(\alpha^2-1\right)-\frac{2{g'}^{(2-\nu)}{L_{m}^{\alpha}(-g)}^2}{c\left({L_{m}^{\alpha-1}(-g)}\right)^2}\nonumber\\&\quad+\frac{{g'}^{(2-\nu)}{L_{m-2}^{\alpha+1}(-g)}}{c\left({L_{m}^{\alpha-1}(-g)}\right)}-\frac{{g'}^{(2-\nu)}{L_{m-1}^{\alpha}(-g)}}{c\left({L_{m}^{\alpha-1}(-g)}\right)}-\frac{{g'}^{(2-\nu)}{L_{m-1}^{\alpha+1}(-g)}}{\alpha c\left({L_{m}^{\alpha-1}(-g)}\right)}+\frac{{g'}^{(2-\nu)}{L_{m-1}^{\alpha+1}(-g)}}{c\left({L_{m}^{\alpha-1}(-g)}\right)}\nonumber\\&\quad-\frac{({\nu}^2-\nu) {g''}^2}{2c{g'}^{(\nu+2)}}-\frac{\nu g'''}{2c{g'}^{(\nu+1)}}+\frac{3\nu^2{g''}^2}{4c{g'}^{(\nu+2)}}+\frac{g'''}{2c{g'}^{(\nu+1)}}-\frac{3{g''}^2}{4c{g'}^{(\nu+2)}}
\end{align}

By following a similar approach as in the case $1$, we observe that the choice  ${g'}^{(2-\nu)} = \kappa \, c \, g $, where $\kappa$ is a constant, will provide a constant term on the right-hand side of the equation, corresponding to the energy term $E$ on the left-hand side. To get increasing eigenvalues for successive $n$ values, it is necessary to restrict the constant $\kappa$ to positive values.The solution of mentioned first order differential equation  ${g'}^{(2-\nu)} = \kappa \, c \, g $ reads $g(x)=x^{\frac{2-\nu}{1-\nu}}$, which generate $M(x)={\left(\frac{2-\nu}{1-\nu}\right)}^2x^{\frac{\nu}{1-\nu}}$. For simplicity we may take $\kappa \, c={(\frac{2-\nu}{1-\nu})}^2$ such that $\kappa = 1$ and $c = {(\frac{2-\nu}{1-\nu})}^2$. Substituting these values in Eq.(\ref{XmEVeff}) we obtain the effective potential and energy eigenvalues as 
\begin{align}\label{Vsm}
 V_{eff}&=\frac{1}{4}\left( x^{\frac{\nu-2}{1-\nu}}\left(\alpha^2-1\right)+ x^{\frac{2-\nu}{1-\nu}}\right)+\Bigg(\frac{2(L_{m-1}^{\alpha}(-x^{\frac{2-\nu}{1-\nu}}))^2}{({L_{m}^{\alpha-1}(-x^{\frac{2-\nu}{1-\nu}})})^2}-\frac{{L_{m-2}^{\alpha+1}(-x^{\frac{2-\nu}{1-\nu}})}}{{L_{m}^{\alpha-1}(-x^{\frac{2-\nu}{1-\nu}})}} +\frac{{L_{m-1}^{\alpha+1}(-x^{\frac{2-\nu}{1-\nu}})}}{\alpha{L_{m}^{\alpha-1}(-x^{\frac{2-\nu}{1-\nu}})}}\nonumber\\&\quad+\frac{{L_{m-1}^{\alpha}(-x^{\frac{2-\nu}{1-\nu}})}}{{L_{m}^{\alpha-1}(-x^{\frac{2-\nu}{1-\nu}})}} -\frac{{L_{m-1}^{\alpha+1}(-x^{\frac{2-\nu}{1-\nu}})}}{{L_{m}^{\alpha-1}(-x^{\frac{2-\nu}{1-\nu}})}}\Bigg)x^{\frac{2-\nu}{1-\nu}}+\frac{\left(1-\nu\right)\left(3-\nu\right)}{4{\left(2-\nu\right)}^2}x^{\frac{\nu-2}{1-\nu}}+V^{II}_c 
\end{align}

\begin{equation}
	E_n^m=\left(n+\frac{\alpha+1}{2}+\frac{m}{\alpha}\right)+V^{II}_c,  ~~~n=0,1,2,...
\end{equation}
where, $V^{II}_c $ is arbitrary constant.The eigenfunctions are obtained using Eqs.(\ref{psi f(x)F(g(x))}) and (\ref{fx})
\begin{equation}
	\psi_n^m(x) = \mathcal{N} \frac{\exp\left(\frac{-x^{\frac{2-\nu}{1-\nu}}}{2}\right)}{x^{\frac{1}{2}}\left(L_{m}^{\alpha-1}(-x^{\frac{2-\nu}{1-\nu}})\right)}x^{\frac{(2-\nu)(1+\alpha)}{2(1-\nu)}}\hat{L}_{n+m,m}^\alpha(x^{\frac{2-\nu}{1-\nu}}),  ~~~n=0,1,2,...
\end{equation}
where $\mathcal{N}$ is the normalization constant and is given by
\begin{equation*}
    \mathcal{N}=\left(\frac{n!}{(n+m+\alpha)\Gamma (n+\alpha)}\right)^{\frac{1}{2}}
\end{equation*}
\begin{figure}[H]
	\centering
	\begin{subfigure}[b]{0.49\textwidth}
		\centering
		\includegraphics[width=0.95 \textwidth]{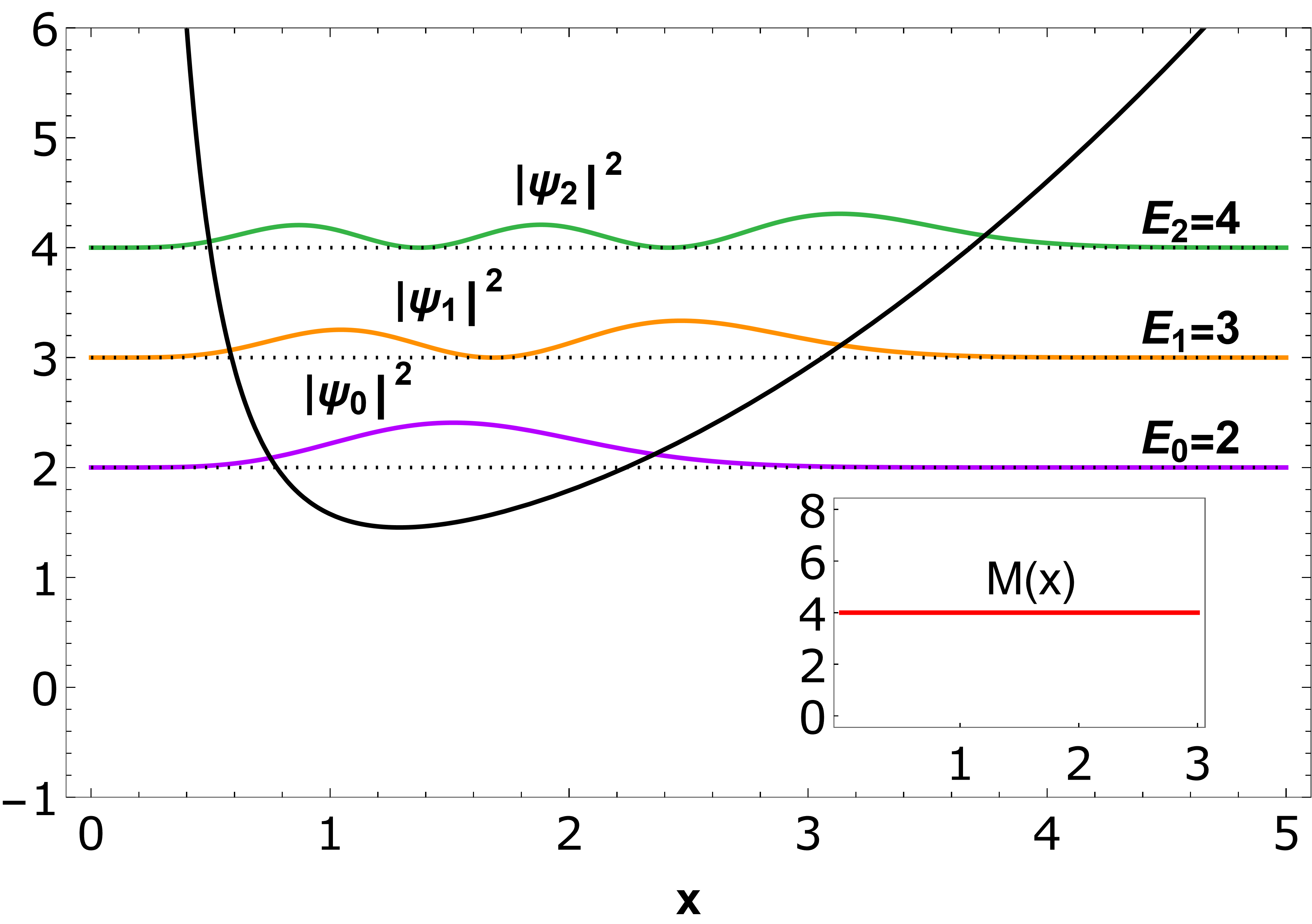}
		\caption{$m=1,\nu=0$}
	\end{subfigure}
	\begin{subfigure}[b]{0.49\textwidth}
		\centering
		\includegraphics[width=0.95 \textwidth]{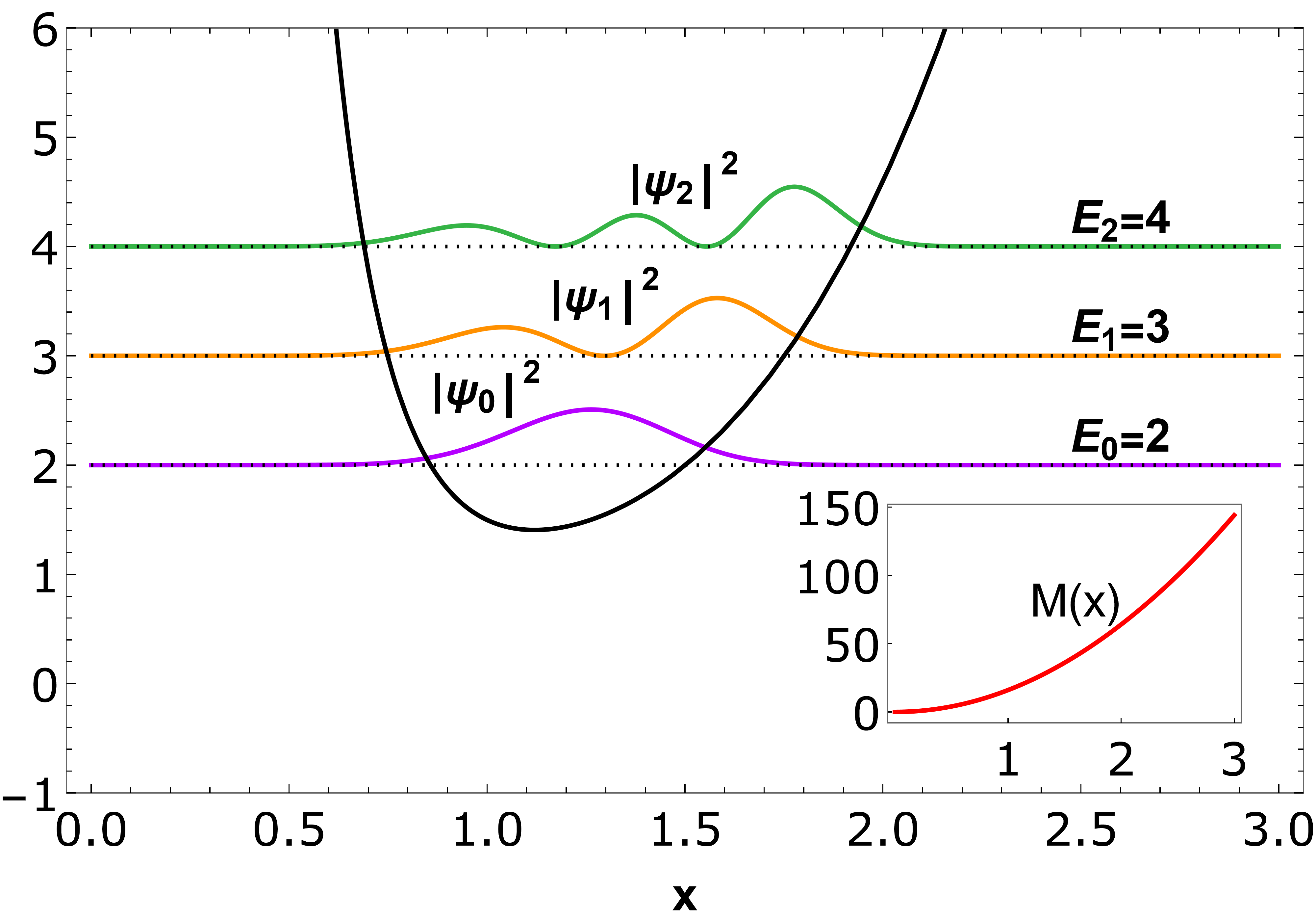}
		\caption{$m=1, \nu=\frac{2}{3}$}
	\end{subfigure}
	\begin{subfigure}[b]{0.49\textwidth}
		\centering
		\includegraphics[width=0.95 \textwidth]{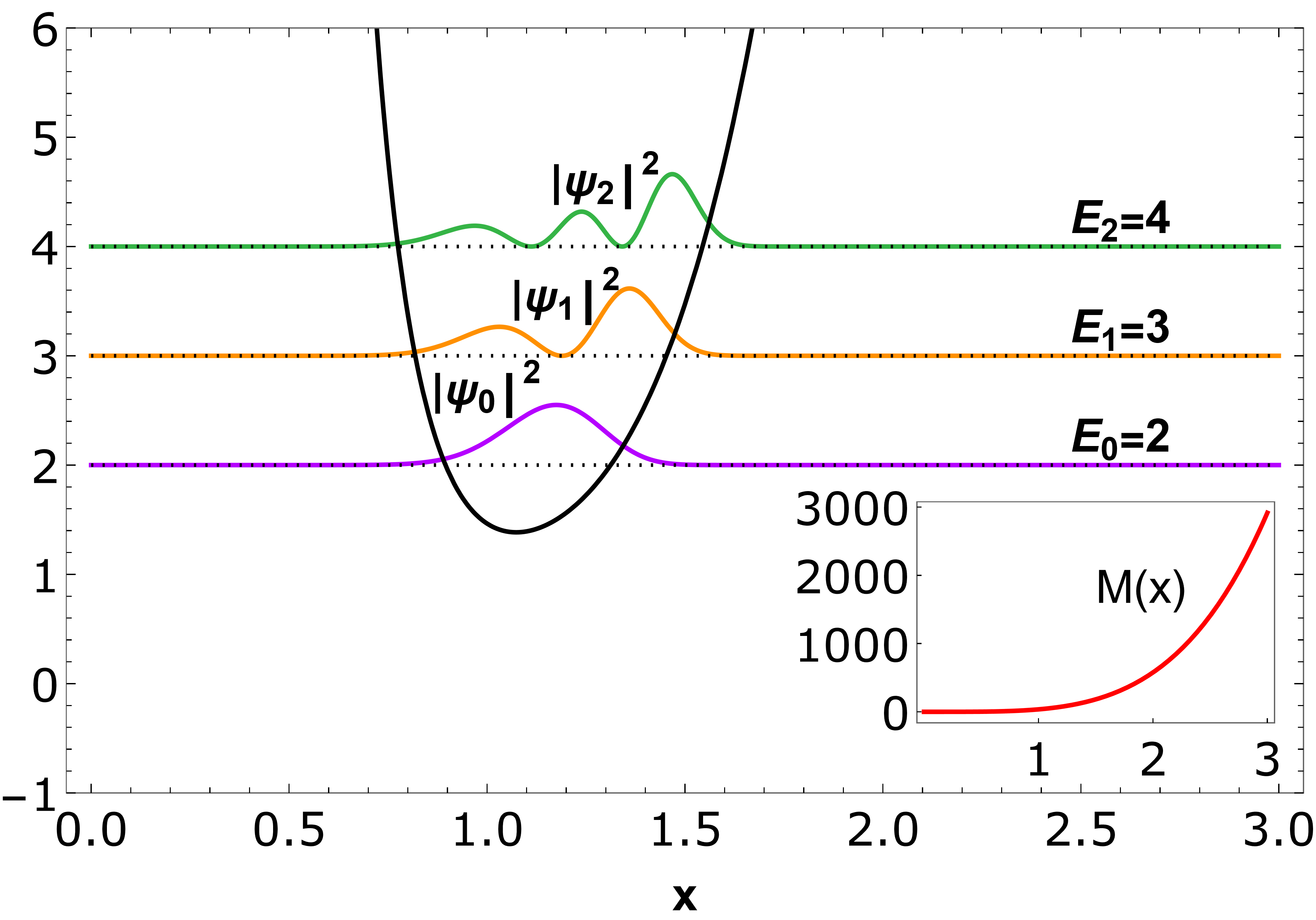}
		\caption{$m=1, \nu=\frac{4}{5}$}
	\end{subfigure}
	\begin{subfigure}[b]{0.49\textwidth}
		\centering
		\includegraphics[width=0.95 \textwidth]{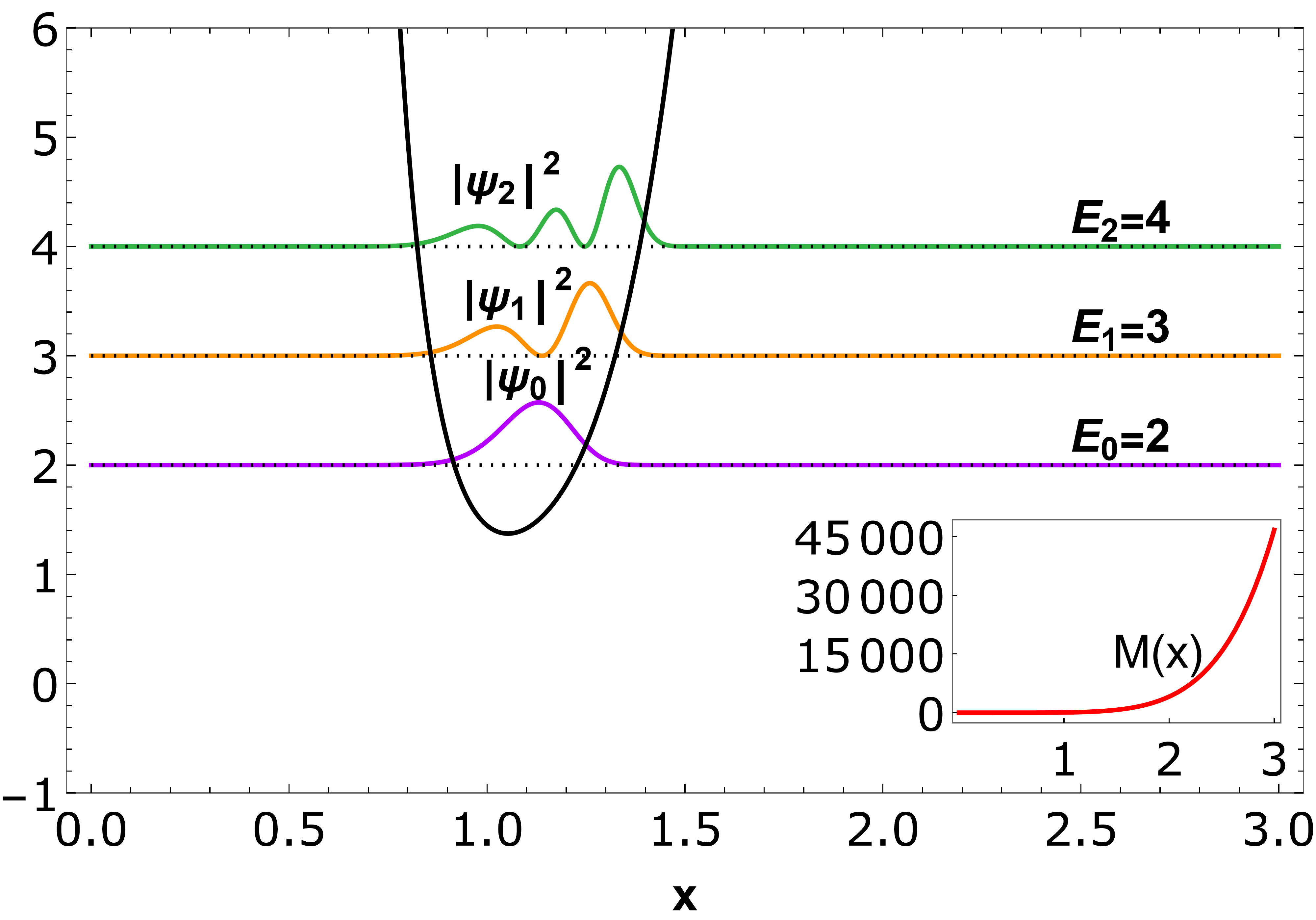}
		\caption{$m=1, \nu=\frac{6}{7}$}
	\end{subfigure}
	\caption{Plot of the potential $V_{eff}$ given in Eq. (\ref{Vsm}) for $m=1$ and different $\nu$ values, square of first three bound state wavefunctions ${|\psi_0|}^2$ (purple line), ${|\psi_1|}^2$ (orange line),  ${|\psi_2|}^2$ (green line) corresponding to different $\nu$ values, for mass function $M(x)$ (red line). We have consider here $\alpha=2$.}
	\label{fig:enter-label1}
\end{figure}

\begin{figure}[H]
	\centering
	\begin{subfigure}[b]{0.49\textwidth}
		\centering
		\includegraphics[width=0.95\textwidth]{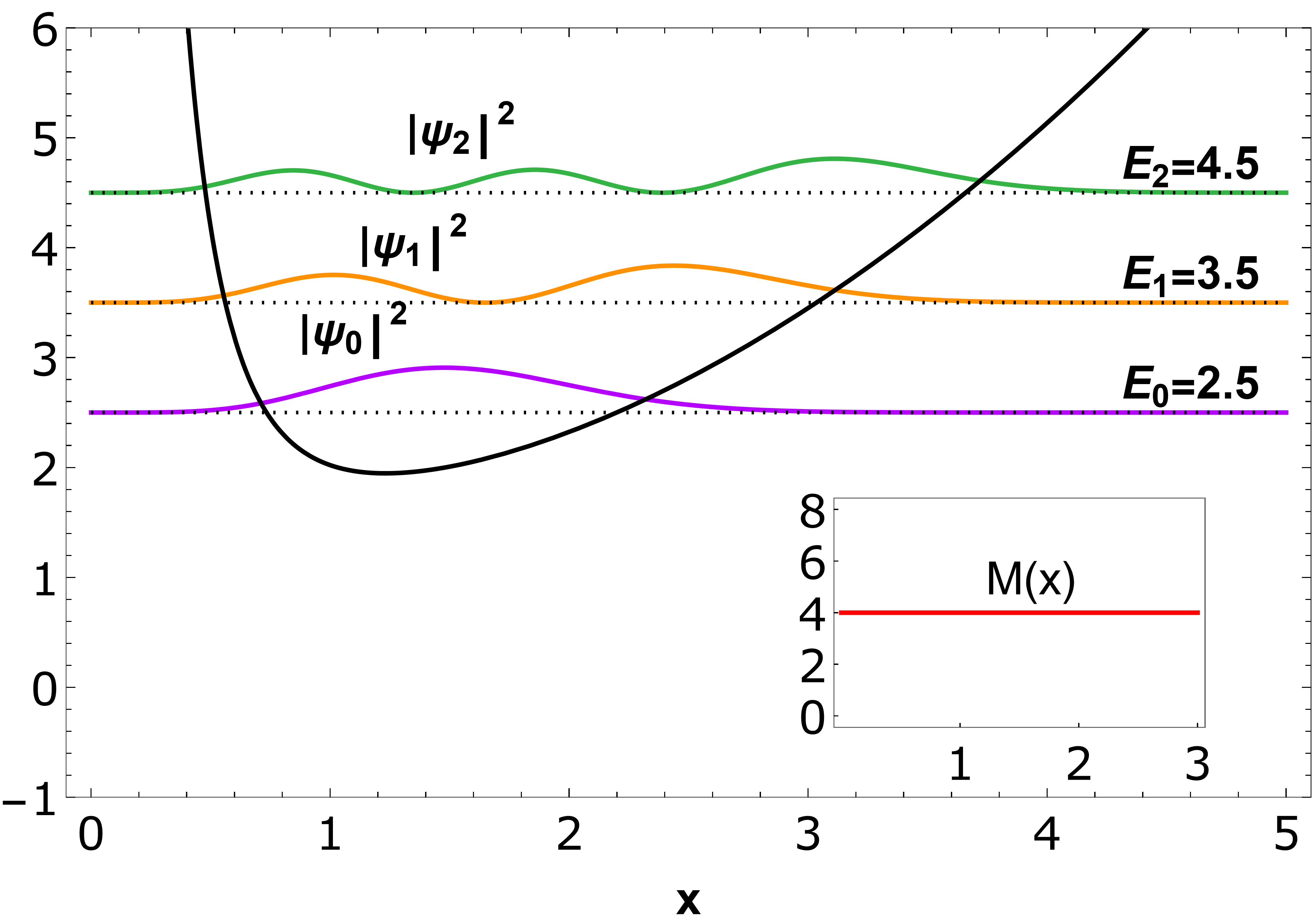}
		\caption{$m=2,\nu=0$}
	\end{subfigure}
	\begin{subfigure}[b]{0.49\textwidth}
		\centering
		\includegraphics[width=0.95\textwidth]{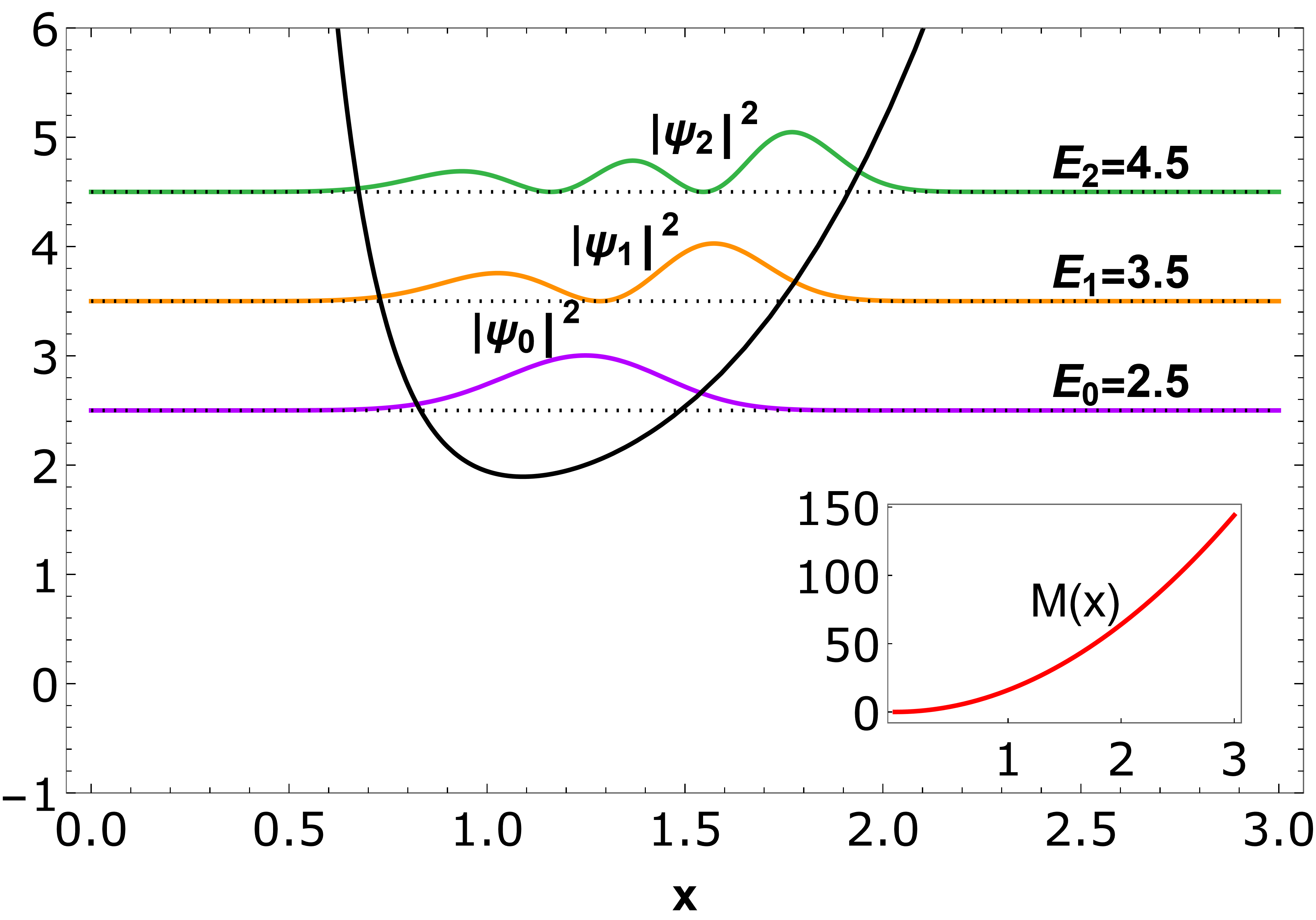}
		\caption{$m=2, \nu=\frac{2}{3}$}
	\end{subfigure}
	\begin{subfigure}[b]{0.49\textwidth}
		\centering
		\includegraphics[width=0.95\textwidth]{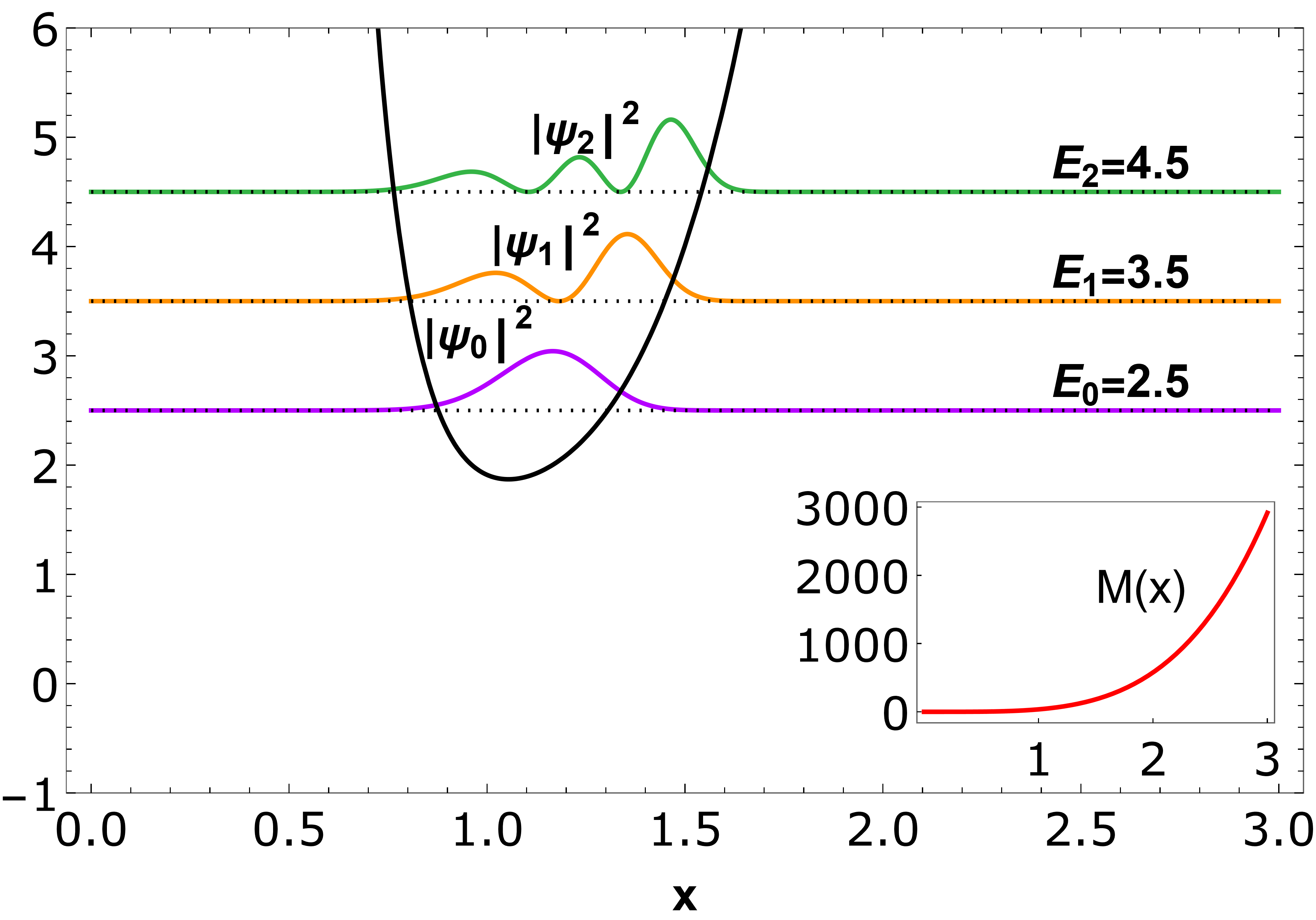}
		\caption{$m=2, \nu=\frac{4}{5}$}
	\end{subfigure}
	\begin{subfigure}[b]{0.49\textwidth}
		\centering
		\includegraphics[width=0.95\textwidth]{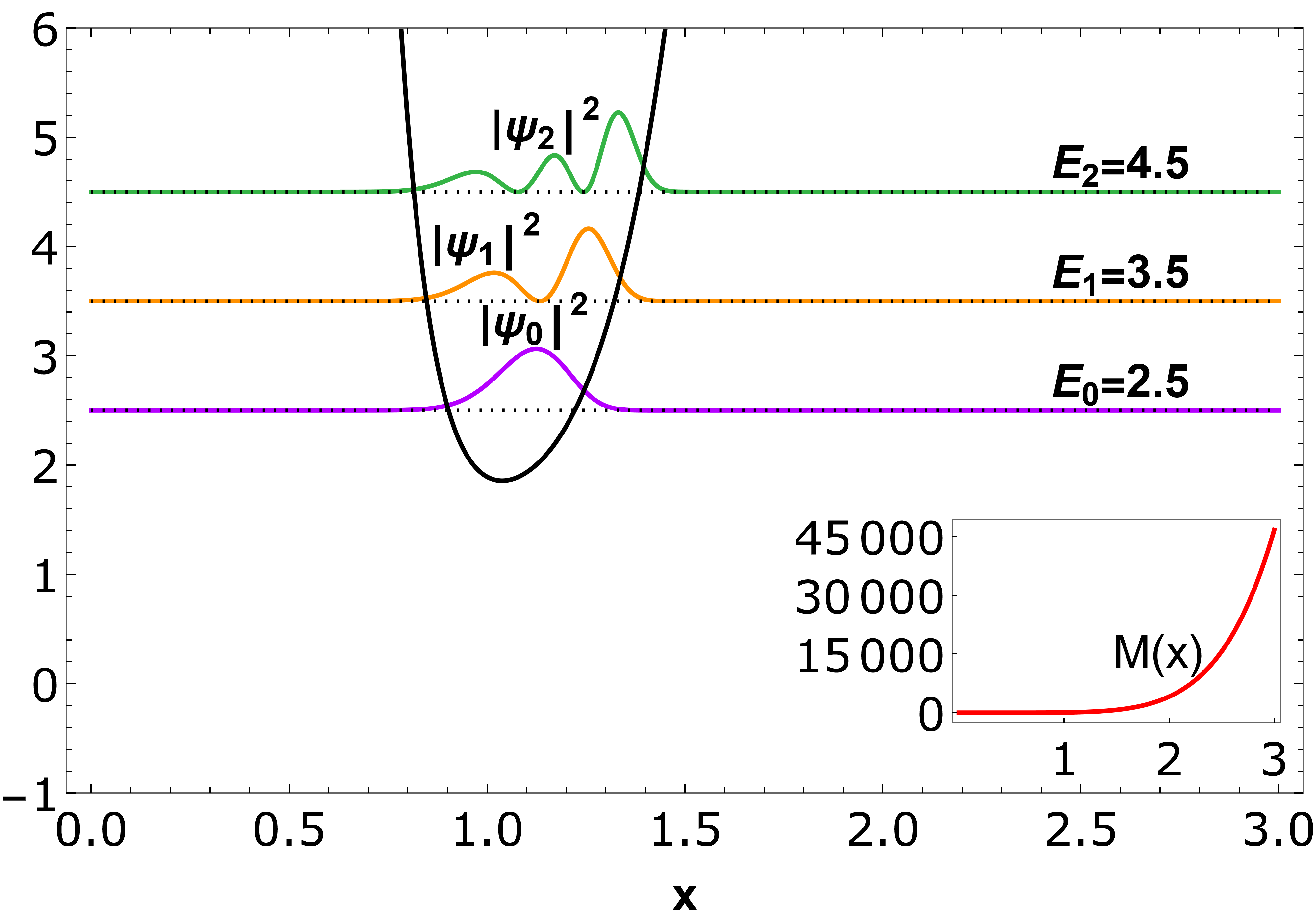}
		\caption{$m=2, \nu=\frac{6}{7}$}
	\end{subfigure}
	\caption{Plot of the potential $V_{eff}$ given in Eq. (\ref{Vsm}) for $m=2$ and different $\nu$ values, square of first three bound state wavefunctions ${|\psi_0|}^2$ (purple line), ${|\psi_1|}^2$ (orange line),  ${|\psi_2|}^2$ (green line) corresponding to different $\nu$ values, for mass function $M(x)$ (red line). We have consider here $\alpha=2$}
	\label{fig:enter-label2}
\end{figure}
\begin{figure}[H]
	\centering
	\begin{subfigure}[b]{0.49\textwidth}
		\centering
		\includegraphics[width=0.95\textwidth]{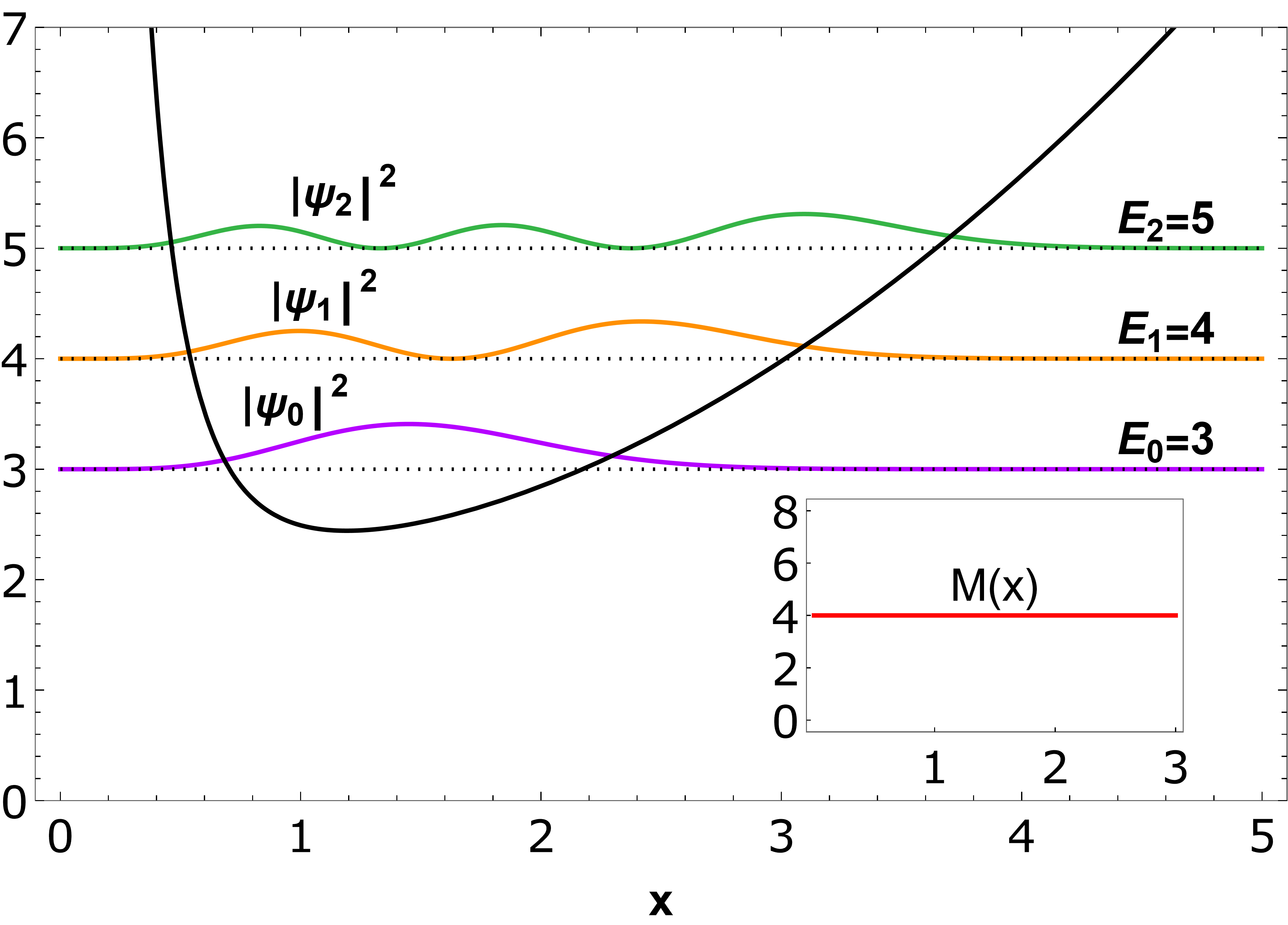}
		\caption{$m=3,\nu=0$}
	\end{subfigure}
	\begin{subfigure}[b]{0.49\textwidth}
		\centering
		\includegraphics[width=0.95\textwidth]{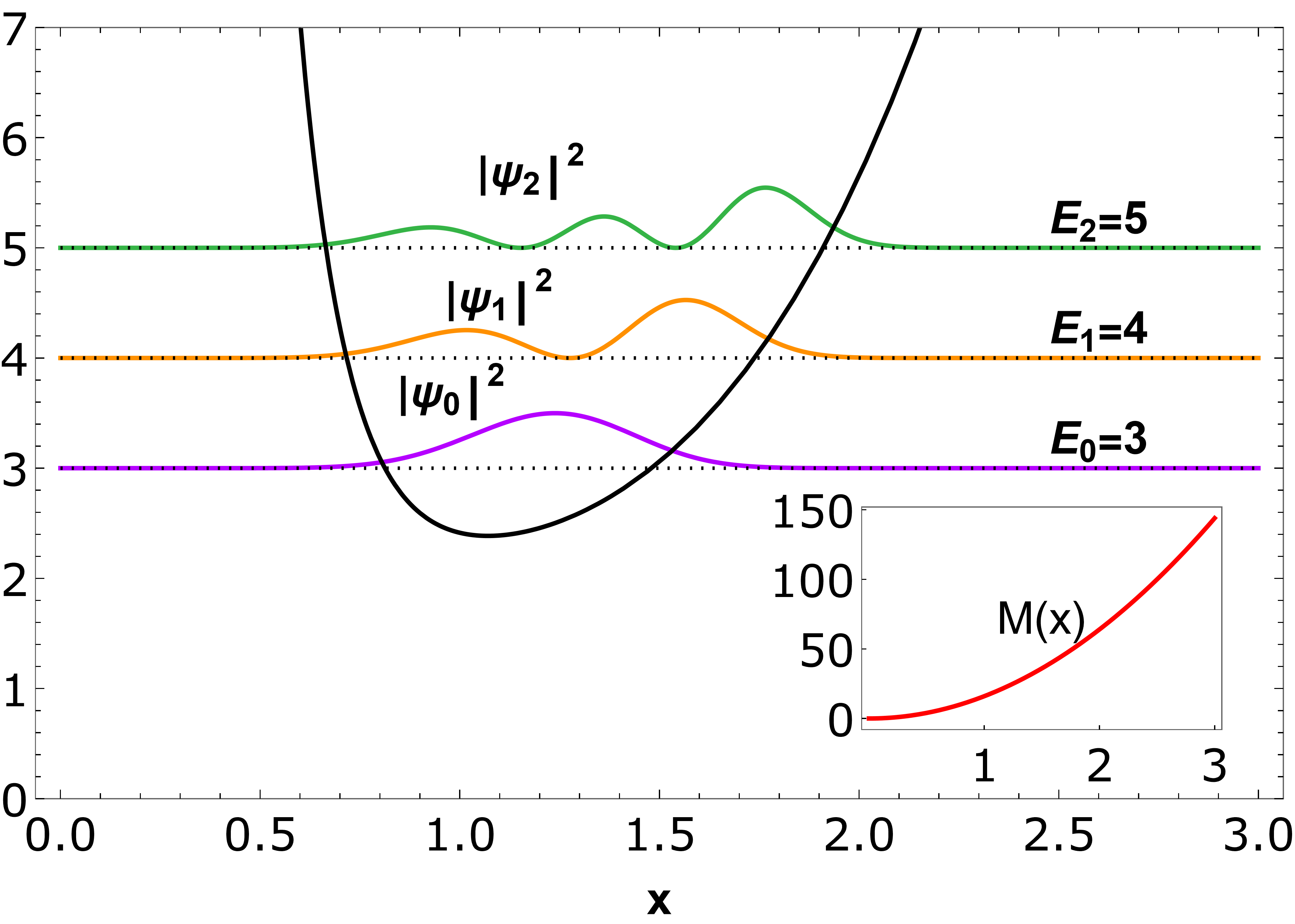}
		\caption{$m=3, \nu=\frac{2}{3}$}
	\end{subfigure}
	\begin{subfigure}[b]{0.49\textwidth}
		\centering
		\includegraphics[width=0.95\textwidth]{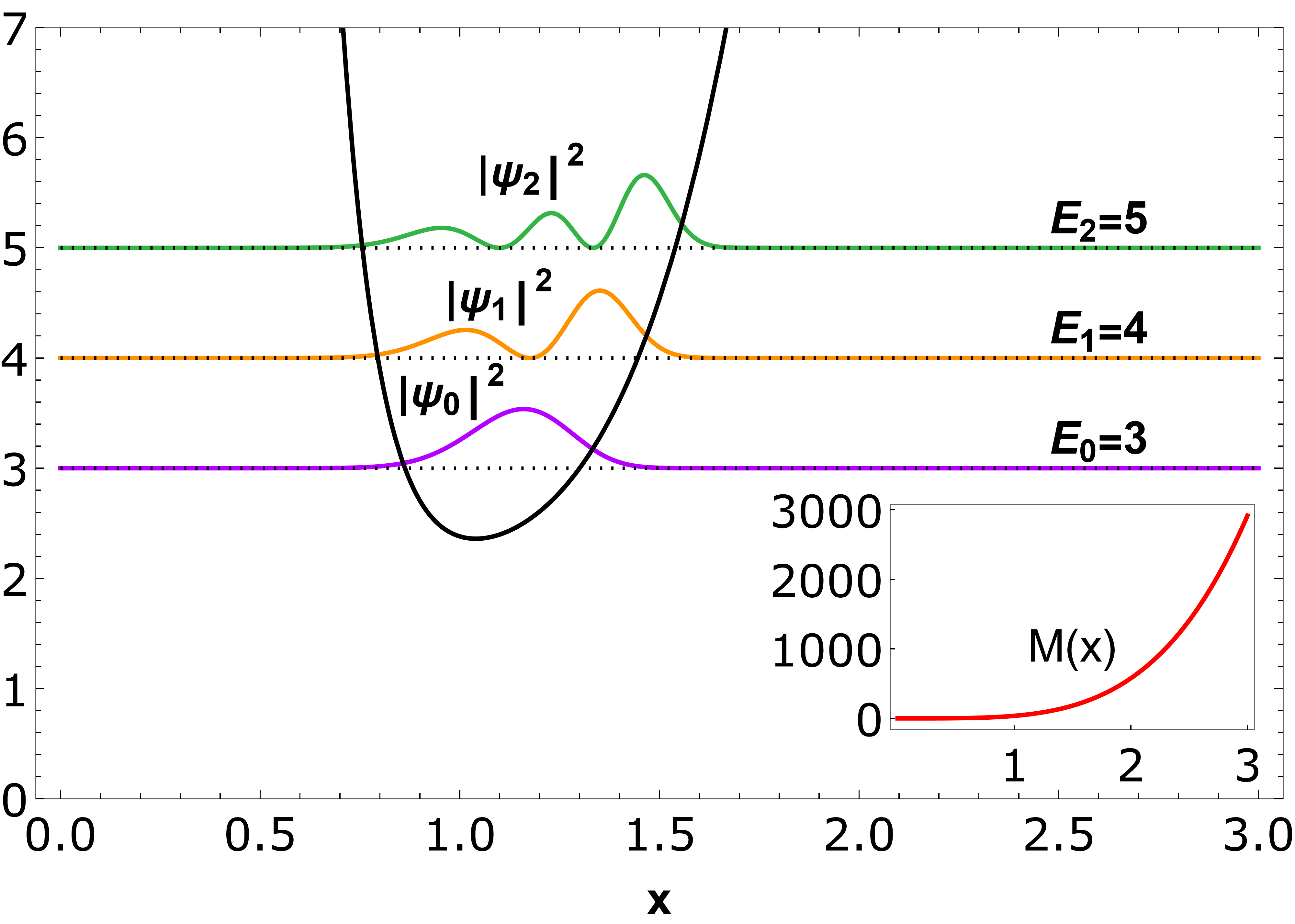}
		\caption{$m=3, \nu=\frac{4}{5}$}
	\end{subfigure}
	\begin{subfigure}[b]{0.49\textwidth}
		\centering
		\includegraphics[width=0.95\textwidth]{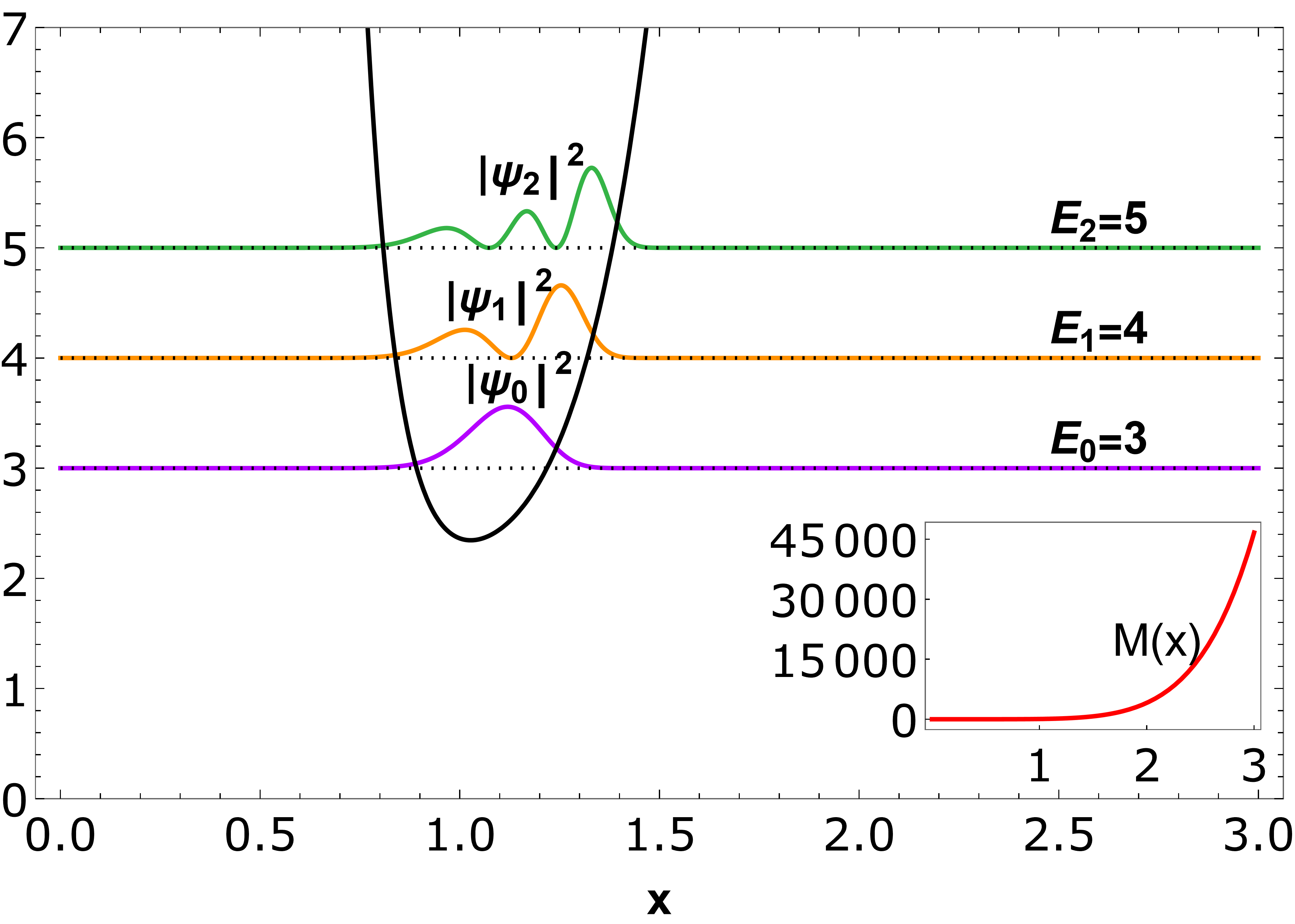}
		\caption{$m=3, \nu=\frac{6}{7}$}
	\end{subfigure}
	\caption{Plot of the potential $V_{eff}$ given in Eq. (\ref{Vsm}) for $m=3$ and different $\nu$ values, square of first three bound state wavefunctions ${|\psi_0|}^2$ (purple line), ${|\psi_1|}^2$ (orange line),  ${|\psi_2|}^2$ (green line) corresponding to different $\nu$ values, for mass function $M(x)$ (red line). We have considered here $\alpha=2$}
	\label{fig:enter-label3}
\end{figure}

\begin{figure}[H]
	\centering
	\begin{subfigure}[b]{0.3\textwidth}
		\centering
		\includegraphics[width=4.7cm]{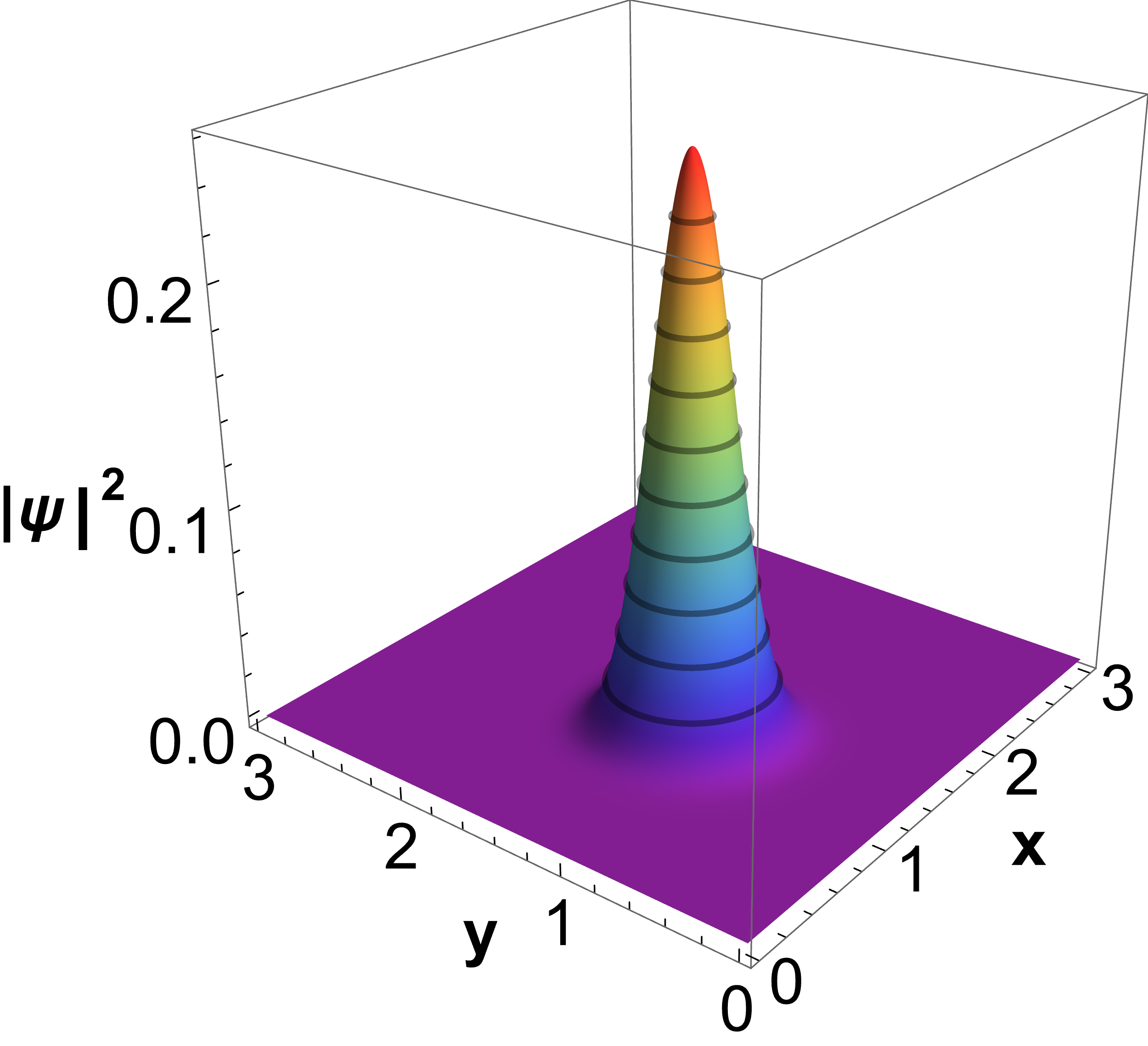}
		\caption{$n_{1}=0,n_{2}=0$}
	\end{subfigure}
	\begin{subfigure}[b]{0.3\textwidth}
		\centering
		\includegraphics[width=4.7cm]{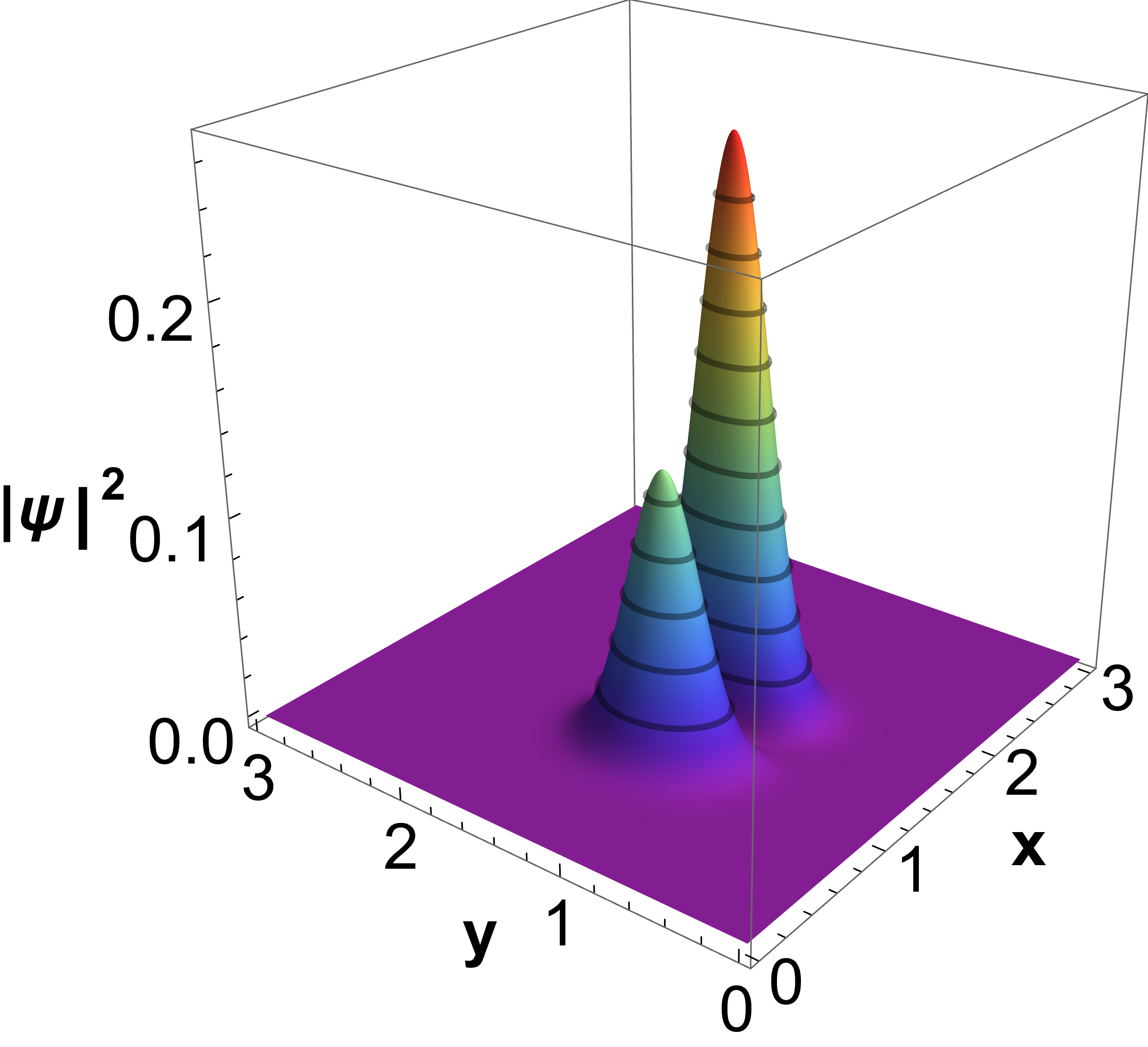}
		\caption{$n_{1}=1,n_{2}=0$}
	\end{subfigure}
	\begin{subfigure}[b]{0.3\textwidth}
		\centering
		\includegraphics[width=4.7cm]{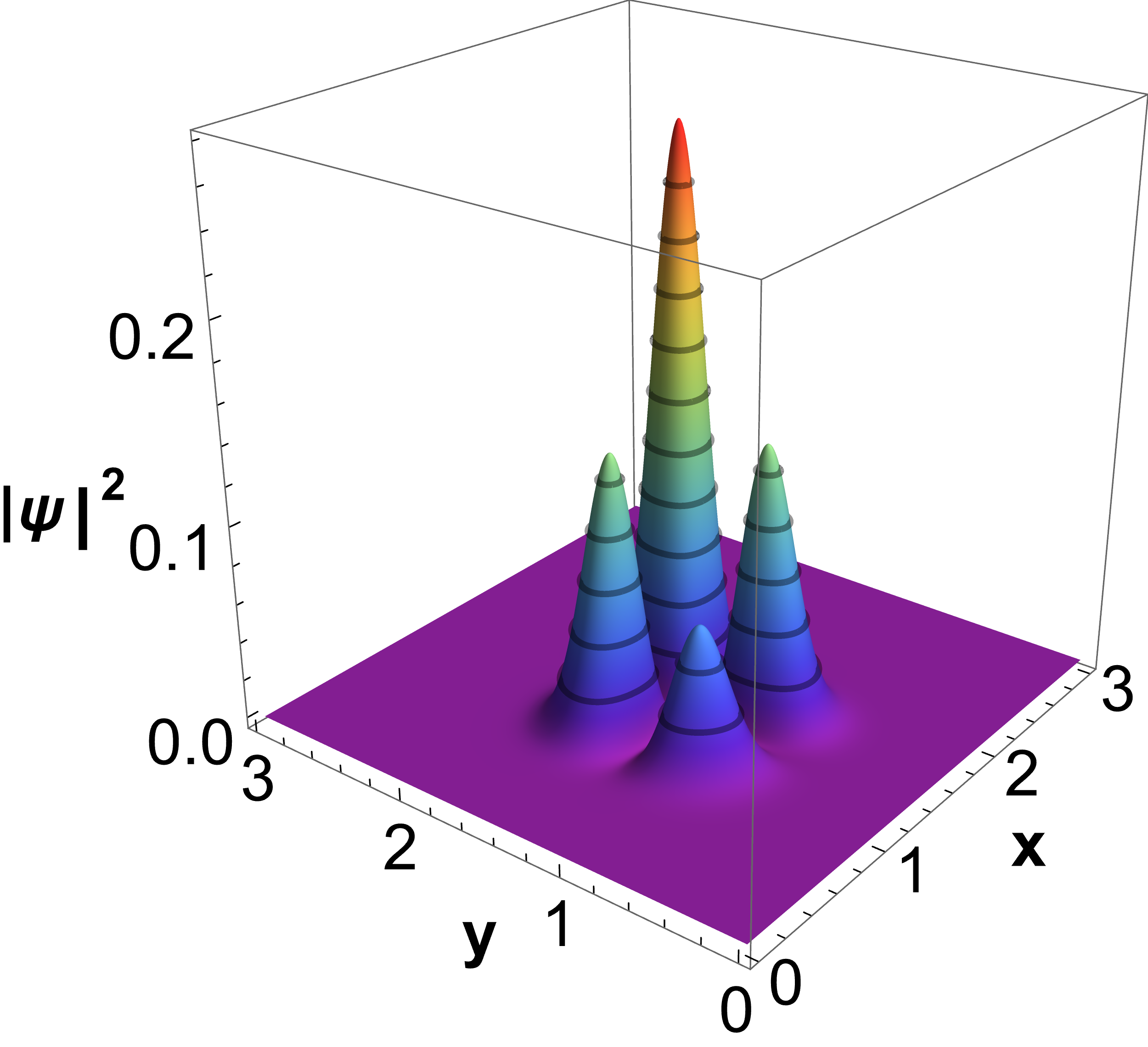}
		\caption{$n_{1}=1,n_{2}=1$}
	\end{subfigure}
	\begin{subfigure}[b]{0.3\textwidth}
		\centering
		\includegraphics[width=4.7cm]{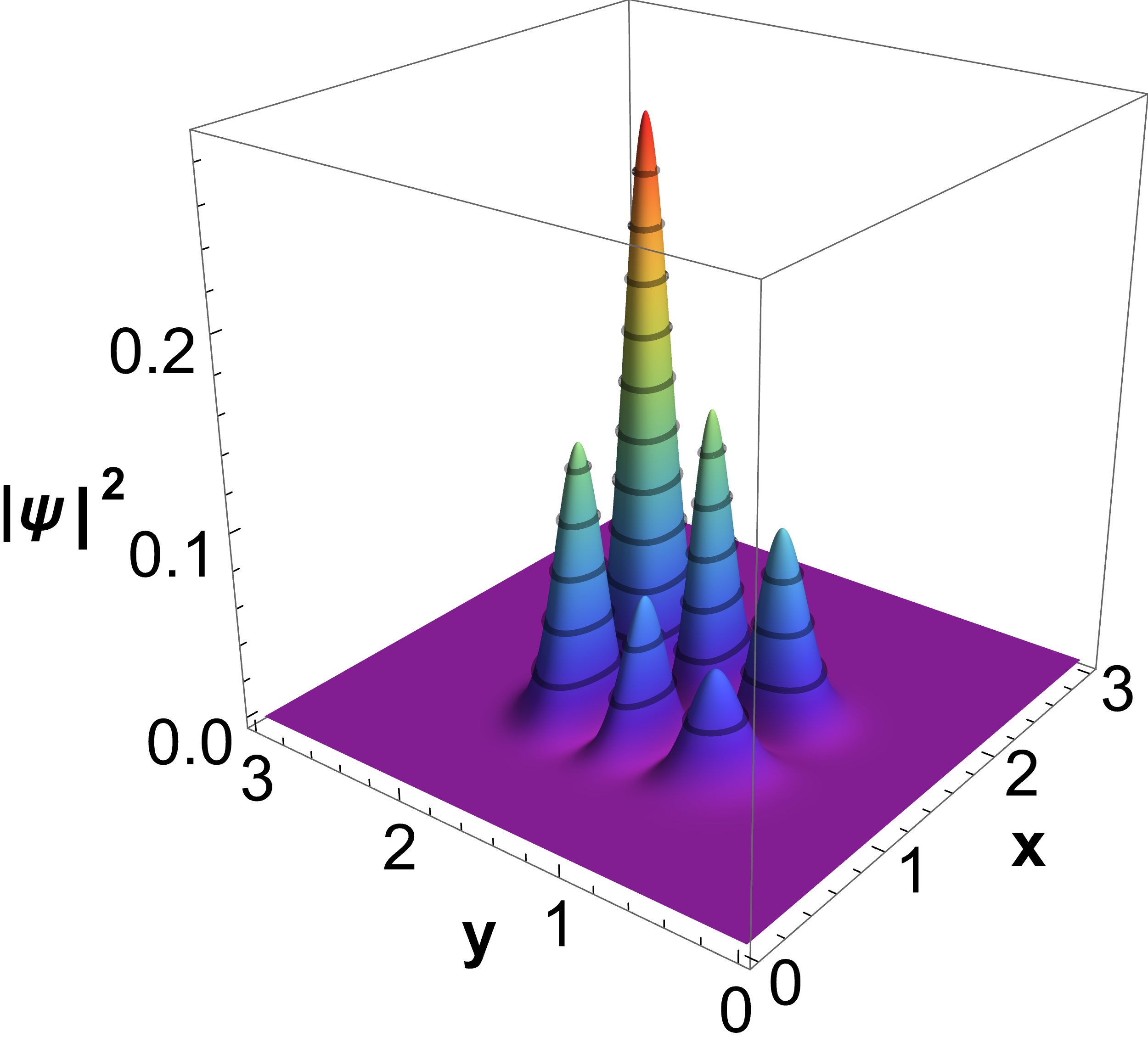}
		\caption{$n_{1}=1,n_{2}=2$}
	\end{subfigure}
	\begin{subfigure}[b]{0.3\textwidth}
		\centering
		\includegraphics[width=4.7cm]{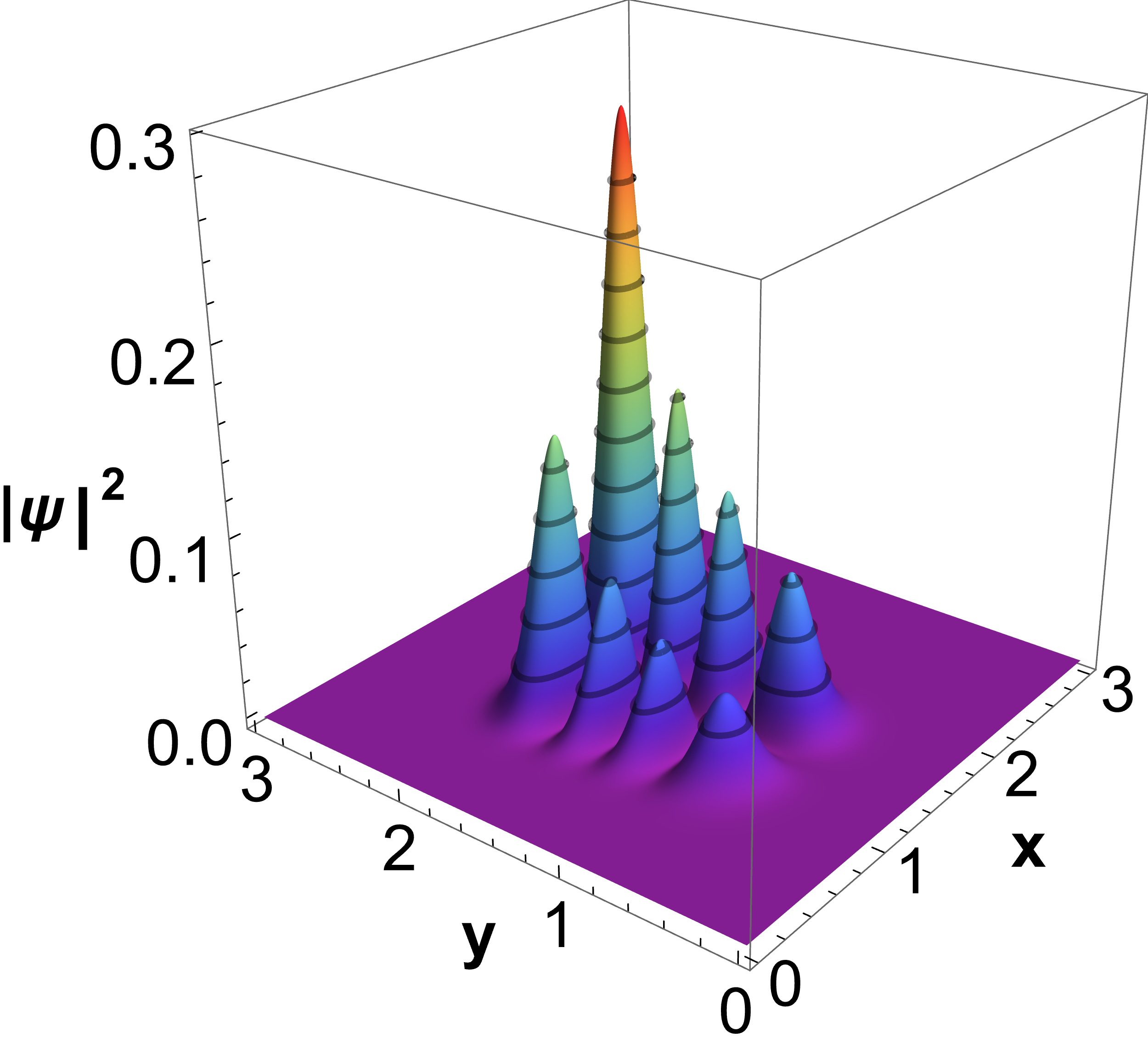}
		\caption{$n_{1}=1,n_{2}=3$}
	\end{subfigure}
	\begin{subfigure}[b]{0.3\textwidth}
		\centering
		\includegraphics[width=4.7cm]{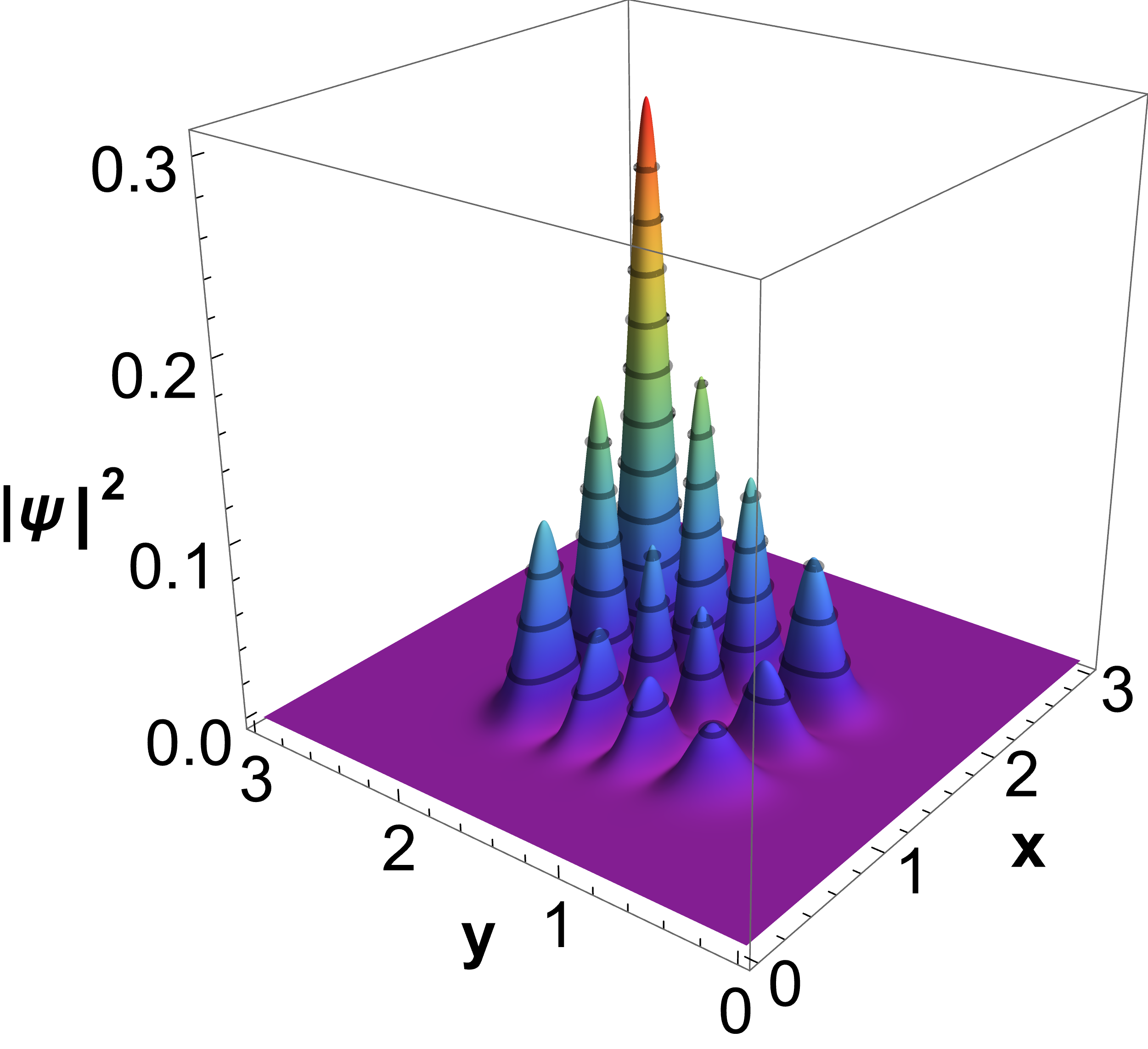}
		\caption{$n_{1}=2,n_{2}=3$}
	\end{subfigure}
	\caption{Plot for the probability density of a two-dimensional potential for $m=1, \nu=\frac{2}{3}$ and different combinations of $n_1$ and $n_2$ values. The wavefunctions are normalized.}
	\label{fig:enter-label4}
\end{figure}

We further generalise this analysis for the two dimensional case. The probability density of a two-dimensional potential is shown in the Fig. \ref{fig:enter-label4}.

\section{Supersymmetric approach}

Within the context of supersymmetric quantum mechanics applied to systems with an effective mass, one can define the lowering and raising operator \cite{plastino1999supersymmetric}
\begin{align}
	\mathcal{A}\psi=\frac{1}{\sqrt{M}}\frac{d\psi}{dx}+\mathcal{W}\psi, \quad  \text{and } \quad \mathcal{A}^\dagger\psi=-\frac{d}{dx}\left(\frac{\psi}{\sqrt{M}}\right)+\mathcal{W}\psi     
\end{align}

where $ \mathcal{W} = -\frac{1}{\sqrt{M}}\frac{\psi_0^{'}}{\psi_0} $ is the superpotential, and $\psi_0$ represents the ground state. The Hamiltonian for a system with an effective mass of (\ref{1}) can be factorized as follows.
\begin{equation}
	H_{eff}=\mathcal{A}^\dagger\mathcal{A}=-\frac{d}{dx}\left(\frac{1}{M(x)}\right)+V_{eff}
\end{equation}
The associated supersymmetric partner Hamiltonian is given by 
\begin{equation}
	H_{eff}^p=\mathcal{A}\mathcal{A}^\dagger=-\frac{d}{dx}\left(\frac{1}{M(x)}\right)+V_{eff}^p
\end{equation}
We see that both partner Hamiltonians $H_{eff}$ and $H_{eff}^p$ describe particles with identical spatial dependence on effective mass but in different potentials $V_{eff}$ and $V_{eff}^p$ respectively. These supersymmetric partner potentials are given by 
\begin{equation}\label{V1}
	V_{eff}=-\left(\frac{\mathcal{W}}{\sqrt{M}}\right)^{'}+\mathcal{W}^2
\end{equation} 
\begin{equation}\label{V2}
	V_{eff}^p= V_{eff}+\frac{2\mathcal{W}^{'}}{\sqrt{M}}-\left(\frac{1}{\sqrt{M}}\right)\left(\frac{1}{\sqrt{M}}\right)^{''}
\end{equation} 
If these potentials are shape invariant, they must satisfy the condition
\begin{equation}\label{SI}
	V_{eff}^p(x,a_1)= V_{eff}(x,a_2)+R(a_1)
\end{equation}
where $a_1$ is a set of parameters, $a_2$ is some function of $a_1$ and $R(a_1)$ is independent of x. For unbroken supersymmetry, the eigenvalues and the eigenfunctions of the two such Hamiltonians with position-dependent effective mass are related by $(n=0,1,2,...)$
\begin{equation}
	E_n^{eff,p}=E_{n+1}^{eff}, ~~~~~~ E_0^{eff}=0
\end{equation}
\begin{equation}\label{wf}
	\psi_n^{eff,p}=[E_{n+1}^{eff}]^{-\frac{1}{2}}\mathcal{A}\psi_{n+1}^{eff}
\end{equation}
\begin{equation}
	\psi_{n+1}^{eff}=[E_{n}^{eff,1}]^{-\frac{1}{2}}\mathcal{A}^\dagger\psi_n^{eff,1}
\end{equation}
The superpotential, corresponding to the potentials derived in case $1$  and case $2$ in the previous section, can be expressed as
\begin{equation}
	\mathcal{W}^{I}=\frac{b}{2}\left((1+\alpha)e^{\frac{bx}{2}}-e^{\frac{-bx}{2}}\left(1+2\left(\frac{L_{m-1}^{\alpha}(-e^{-bx})}{{L_{m}^{\alpha-1}(-e^{-bx})}}-\frac{L_{m-1}^{\alpha+1}(-e^{-bx})}{{L_{m}^{\alpha}(-e^{-bx})}}\right)\right)\right) 
\end{equation}
and
\begin{align}
	\mathcal{W}^{II} = \left( \frac{1}{2} - \left(\frac{L_{m-1}^{\alpha +1}\left(-x^l\right)}{L_m^{\alpha }\left(-x^l\right)}-\frac{L_{m-1}^{\alpha }\left(-x^l\right)}{L_m^{\alpha -1}\left(-x^l\right)}\right) \right) x^{\frac{l}{2}} + \frac{1- l(\alpha+1)}{2 l} x^{-\frac{l}{2}}
\end{align}
respectively, where $l=\frac{2-\nu}{1-\nu}$.\\
From the superpotential $\mathcal{W}^I$ (for case $1$), the partner potentials can be derived using equations (\ref{V1}) and (\ref{V2})
\begin{align}\label{VS}
 V_{eff}^I&=\frac{b^2}{4}\left( e^{bx}\left(\alpha^2-1\right)+ e^{-bx}\right) + b^2\Bigg(\frac{2{L_{m-1}^{\alpha}(-e^{-bx})}^2}{({L_{m}^{\alpha-1}(-e^{-bx})})^2}-\frac{{L_{m-2}^{\alpha+1}(-e^{-bx})}}{{L_{m}^{\alpha-1}(-e^{-bx})}}+\frac{{L_{m-1}^{\alpha+1}(-e^{-bx})}}{\alpha{L_{m}^{\alpha-1}(-e^{-bx})}}\nonumber\\&\quad+\frac{{L_{m-1}^{\alpha}(-e^{-bx})}}{{L_{m}^{\alpha-1}(-e^{-bx})}}-\frac{{L_{m-1}^{\alpha+1}(-e^{-bx})}}{{L_{m}^{\alpha-1}(-e^{-bx})}}\Bigg)e^{-bx} -b^2\left(\frac{2m+\alpha+\alpha^2}{2\alpha}\right)
\end{align}

\begin{align}\label{VS1}
 V_{eff}^{I,p}&=\frac{b^2}{4}\left( e^{bx}\alpha\left(\alpha+2\right)+ e^{-bx}\right) + b^2\Bigg(\frac{2{L_{m-1}^{\alpha+1}(-e^{-bx})}^2}{({L_{m}^{\alpha}(-e^{-bx})})^2}-\frac{{L_{m-2}^{\alpha+2}(-e^{-bx})}}{{L_{m}^{\alpha}(-e^{-bx})}}+\frac{{L_{m-1}^{\alpha+2}(-e^{-bx})}}{\alpha{L_{m}^{\alpha}(-e^{-bx})}}\nonumber\\&\quad+\frac{{L_{m-1}^{\alpha+1}(-e^{-bx})}}{{L_{m}^{\alpha}(-e^{-bx})}}-\frac{{L_{m-1}^{\alpha+2}(-e^{-bx})}}{{L_{m}^{\alpha}(-e^{-bx})}}\Bigg)e^{-bx}-b^2\left(\frac{\alpha+1}{2}+\frac{m}{\alpha+2}\right)
\end{align} 
Also from superpotential $\mathcal{W}^{II}$ (for case $2$) and equations (\ref{V1}) and (\ref{V2}) we can write the partner potentials 
\begin{align}\label{VSM}
 V^{II}_{eff}&=\frac{1}{4}\left( x^{\frac{\nu-2}{1-\nu}}\left(\alpha^2-1\right)+ x^{\frac{2-\nu}{1-\nu}}\right)+\Bigg(\frac{2{L_{m-1}^{\alpha}(-x^{\frac{2-\nu}{1-\nu}})}^2}{({L_{m}^{\alpha-1}(-x^{\frac{2-\nu}{1-\nu}})})^2}-\frac{{L_{m-2}^{\alpha+1}(-x^{\frac{2-\nu}{1-\nu}})}}{{L_{m}^{\alpha-1}(-x^{\frac{2-\nu}{1-\nu}})}}+\frac{{L_{m-1}^{\alpha+1}(-x^{\frac{2-\nu}{1-\nu}})}}{\alpha{L_{m}^{\alpha-1}(-x^{\frac{2-\nu}{1-\nu}})}}\nonumber\\&\quad+\frac{{L_{m-1}^{\alpha}(-x^{\frac{2-\nu}{1-\nu}})}}{{L_{m}^{\alpha-1}(-x^{\frac{2-\nu}{1-\nu}})}} -\frac{{L_{m-1}^{\alpha+1}(-x^{\frac{2-\nu}{1-\nu}})}}{{L_{m}^{\alpha-1}(-x^{\frac{2-\nu}{1-\nu}})}}\Bigg)x^{\frac{2-\nu}{1-\nu}}+\frac{(1-\nu)(3-\nu)}{4{(2-\nu)}^2}x^{\frac{\nu-2}{1-\nu}}-\frac{2m+\alpha+\alpha^2}{2\alpha}
\end{align}
and
\begin{align}\label{VSM1}
 V^{II,p}_{eff}&=\frac{1}{4}\left( x^{\frac{\nu-2}{1-\nu}}\alpha(\alpha+2)+ x^{\frac{2-\nu}{1-\nu}}\right)+\Bigg(\frac{2{L_{m-1}^{(\alpha+1)}(-x^{\frac{2-\nu}{1-\nu}})}^2}{({L_{m}^{\alpha}(-x^{\frac{2-\nu}{1-\nu}})})^2}-\frac{{L_{m-2}^{(\alpha+2)}(-x^{\frac{2-\nu}{1-\nu}})}}{{L_{m}^{\alpha}(-x^{\frac{2-\nu}{1-\nu}})}}+\frac{{L_{m-1}^{(\alpha+2)}(-x^{\frac{2-\nu}{1-\nu}})}}{(\alpha+1){L_{m}^{\alpha}(-x^{\frac{2-\nu}{1-\nu}})}}\nonumber\\&\quad+\frac{{L_{m-1}^{(\alpha+1)}(-x^{\frac{2-\nu}{1-\nu}})}}{{L_{m}^{\alpha}(-x^{\frac{2-\nu}{1-\nu}})}} -\frac{{L_{m-1}^{(\alpha+2)}(-x^{\frac{2-\nu}{1-\nu}})}}{{L_{m}^{\alpha}(-x^{\frac{2-\nu}{1-\nu}})}}\Bigg)x^{\frac{2-\nu}{1-\nu}}+\frac{(1-\nu)(3-\nu)}{4{(2-\nu)}^2}x^{\frac{\nu-2}{1-\nu}}-\left(\frac{\alpha}{2}+\frac{m}{\alpha+1}\right)
\end{align}
The potentials $V^I_{eff}$ and $V^{II}_{eff}$ obtained in Eq.(\ref{VS}) and (\ref{VSM})are same with potentials in Eq. (\ref{Vs}) and (\ref{Vsm}) for $V^I_c=-b^2\left(\frac{2m+\alpha+\alpha^2}{2\alpha}\right)$ and $V^{II}_c=-\frac{2m+\alpha+\alpha^2}{2\alpha}$ respectively.\\
From Eqs. (\ref{VS}) and (\ref{VS1}), we note that the potential and its supersymmetric counterpart satisfy the condition
\begin{equation}
	V^{I,p}_{eff}(x,\alpha)= V^I_{eff}(x,\alpha+1)+b^2
\end{equation}
Also from Eqs. (\ref{VSM}) and (\ref{VSM1}) we see that the partner potentials satisfy the similar relation
\begin{equation}
	V^{II,p}_{eff}(x,\alpha)= V^{II}_{eff}(x,\alpha+1)+1
\end{equation}
It's evident that the potentials obtained for case $1$  and case $2$ satisfy the condition of shape invariant potentials given in Eq. (\ref{SI}). Thus, we deduce that the newly introduced potentials exhibit shape invariance.\\
We've deduced the eigenstates for the partner potential associated with potential (\ref{Vs}) as 
\begin{equation}
	\psi^p_{n,m}(x) = \propto \frac{\exp{[-\frac{1}{2}((\alpha+2)bx+e^{-bx})]}}{L_{m}^{\alpha}(-e^{-bx})}\hat{L}_{n+m,m}^{\alpha+1}(e^{-bx})  ~~~n=0,1,2,...
\end{equation}
and the eigenstate of the partner potential of Eq. (\ref{Vsm}) is given by
\begin{equation}
	\psi^p_{n,m}(x) \propto \frac{\exp\left(\frac{-x^{\frac{2-\nu}{1-\nu}}}{2}\right)}{\left(L_{m}^{\alpha}(-x^{\frac{2-\nu}{1-\nu}})\right)}x^{\frac{(2+\alpha)+(\alpha+1)(1-\nu)}{2(1-\nu)}}\hat{L}_{n+m,m}^{\alpha+1}(x^{\frac{2-\nu}{1-\nu}}),~~~~n=0,1,2,...  
\end{equation}


\section{Results and Discussions}
The study of systems endowed with position-dependent mass (PDM) is a subject of great interest in many branches of physics  due to its utmost relevance in a wide variety of physical situations. However only a few PDM systems are solved  exactly for some specific position dependent of  Mass terms, i.e. $M(x) \propto [g^\prime (x)]^\mu$ for $\mu= 1,2 $ and $-1$.  In this  article we have considered a new type of  position dependent mass term,  which is proportional to $\left( {g'(x)} \right)^\nu $, \ $\nu =\frac{2\eta}{2\eta+1},$ with $\eta= 0,1,2\cdots $   and have
obtained the solutions  of position-dependent mass Schr\"{o}dinger equation in terms of $X_m$ Laguerre polynomials. Furthermore, we have obtained one parameter family of isochronous potentials, which are exactly solvable. The solution corresponding to these potentials are shown to be in terms of $X_m$ Laguerre polynomials. We have shown the SUSY for PDM systems and for this system, we find the one-parameter family of isochronous potentials and its partner potentials. We show the exact solvability of the system is due to the underlying shape invariance of these SUSY potentials.  


\paragraph{}
{\it \bf{Acknowledgments}}:
 BPM acknowledges the support from the research grant under IoE scheme (Number - 6031), BHU, UGC Government of India. SY acknowledges the fruitful discussion with Sudhanshu Shekhar, BHU. RG acknowledges the financial support from IOE, BHU for a short visit to BHU during which some part of the work has been carried out.

\bibliographystyle{elsarticle-num}
\bibliography{Ref}

\end{document}